\documentclass[prd,article,nofootinbib,twocolumn,preprintnumbers]{revtex4}
\usepackage{graphicx,amsfonts,color,comment,amsmath,hyperref,float}
\usepackage{amssymb}	
\def\sec#1{{sec.~(\ref{#1})}}
\usepackage{mathrsfs,amssymb}  
\usepackage{cancel}
\usepackage[normalem]{ulem}

\newcommand{\be}{\begin{equation}}
\newcommand{\ee}{\end{equation}}
\newcommand{\bea}{\begin{eqnarray}}
\newcommand{\eea}{\end{eqnarray}}

\def\fig#1{{Fig.~\ref{#1}}}
\def\eq#1{{Eq.~(\ref{#1})}}

\newcommand{\GeV}{\,\text{GeV}}
\newcommand{\TeV}{\,\text{TeV}}
\newcommand{\missEt}{\cancel{\it{E}}_{T}}
\newcommand{\ifb}{{\rm fb}^{-1}}

\begin{document}

\title{New or $\nu$ Missing Energy?  \\ Discriminating Dark Matter from Neutrino Interactions at the LHC}

\author{Diogo Buarque Franzosi}
\author{Mads T. Frandsen}
\author{Ian M. Shoemaker}
\affiliation{{\color{black} CP$^{3}$-Origins} \& Danish Institute for Advanced Study {\color{black}, Danish IAS}, University of Southern Denmark, Campusvej 55, DK-5230 Odense M, Denmark \\ {\tt franzosi@cp3-origins.net, frandsen@cp3-origins.net, shoemaker@cp3.dias.sdu.dk} \smallskip}

\date{\today}
\begin{abstract}
Missing energy signals such as monojets are a possible signature of Dark Matter (DM) at colliders. However, neutrino interactions beyond the Standard Model may also produce missing energy signals.  In order to conclude that new ``missing particles'' are observed the hypothesis of BSM neutrino interactions must be rejected. In this paper, we first derive new limits on these Non-Standard neutrino Interactions (NSIs) from LHC monojet data. 
For heavy NSI mediators, these limits are much stronger than those coming from traditional low-energy $\nu$ scattering or $\nu$ oscillation experiments for some flavor structures.
Monojet data alone can be used to infer the mass of the ``missing particle'' from the shape of the missing energy distribution. In particular, 13 TeV LHC data will have sensitivity to DM masses greater than $\sim$ 1 TeV.  
In addition to the monojet channel, NSI can be probed in multi-lepton searches which we find to yield stronger limits at heavy mediator masses.  
The sensitivity offered by these multi-lepton channels provide a method to reject or confirm the DM hypothesis in missing energy searches.

\end{abstract}
\preprint{
CP3-Origins-2015-031 DNRF90,
DIAS-2015-31}


\maketitle
\section{Introduction}

Missing energy signals 
are the tell-tale clue of the production of stable neutral objects. Indeed the imbalance of momentum and energy is in fact precisely the way in which the neutrino was first discovered. Supposing that the LHC finds anomalous ``missing energy'' events above SM backgrounds, the determination of its origin will be of paramount importance.  As known sources of missing energy, 
a plausible origin
of new missing energy data will be neutrinos. However, new neutral particles beyond Standard Model (BSM) such as dark matter can also produce missing energy signals at colliders. 

In this paper we explore how LHC data can be used to distinguish these two potential sources of missing energy.
We illustrate that both the DM mass and the $SU(2)$ charge of neutrinos can be used to discriminate between singlet DM and SM neutrinos.~\footnote{If DM itself transforms non-trivially under $SU(2)$ the situation is more complex. We leave for future work a systematic study in this direction but note that some of the implications of $SU(2)$ charged DM in a variety of representations has been studied in e.g.~\cite{Cirelli:2005uq}.}  For simplicity we will focus on the so-called ``monojet'' signature in which a single hard jet recoils against ``nothing''~\cite{Birkedal:2004xn,Cao:2009uw,Beltran:2010ww,Goodman:2010yf,Goodman:2010ku,Bai:2010hh,Fortin:2011hv,Graesser:2011vj,Fox:2011pm,Friedland:2011za,Shoemaker:2011vi,An:2012va,Fox:2012ee,Carpenter:2012rg,Chatrchyan:2012tea,Frandsen:2012rk,Haisch:2012kf,Bell:2012rg,Fox:2012ru,Zhou:2013fla,Busoni:2013lha,An:2013xka,Buchmueller:2013dya,Busoni:2014sya,Buchmueller:2014yoa,Abdallah:2014hon,Jacques:2015zha,Chala:2015ama,Bell:2015sza}. Previous work in this direction using Tevatron and early LHC data was carried out in~\cite{Friedland:2011za}. 

We begin by reviewing current experimental limits on neutrino-proton interactions in the context of effective field theory (EFT).
Up to dimension 6, we can have 
\begin{itemize}
\item{Neutrino magnetic dipole moments}:
\be
\mathscr{L} \supset \mu_{\nu} F^{\mu\nu} \overline{\nu} \sigma_{\mu \nu} \nu,
\ee
where the spin matrix is $\sigma_{\mu \nu} \equiv i \left[ \gamma_{\mu}, \gamma_{\nu}\right]/2$, $\mu_{\nu}$ is the magnetic moment (measured in units of the Bohr magneton $\mu_{B} \equiv e/\left(2m_{e} \right)$, where $e, m_{e}$ are the charge and mass of the electron).
\item{Non-standard neutrino interactions (NSIs)},  
\be
\mathscr{L}_{NSI} =-2\sqrt{2}G_{F}\varepsilon^{fP}_{\alpha \beta}\left(\overline{\nu}_{\alpha}\gamma_{\rho}\nu_{\beta}\right)\left(\overline{f}\gamma^{\rho}Pf \right).
\label{eq:NSI}
\ee
where the matrix $\varepsilon^{fP}_{\alpha \beta}$ specifies the strength of the $\nu$-$f$ interaction, in units of Fermi's constant, $G_{F} \equiv 1/\sqrt{2} v_{EW}^{2} \simeq 1.2 \times 10^{-5}~{\rm GeV}^{-2}$, with $v_{EW} = 246$ GeV. The labels $\alpha,\beta$ are flavor indices running over $e, \mu, \tau$, and $P$ is a projection operator. We take $f$ to be any SM fermion (though only the {\it vector} components of $f = e,u,d$ are relevant for neutrino oscillations). 

\end{itemize}
%

\begin{figure*}[t] 
\begin{center}
 \includegraphics[width=.25\textwidth]{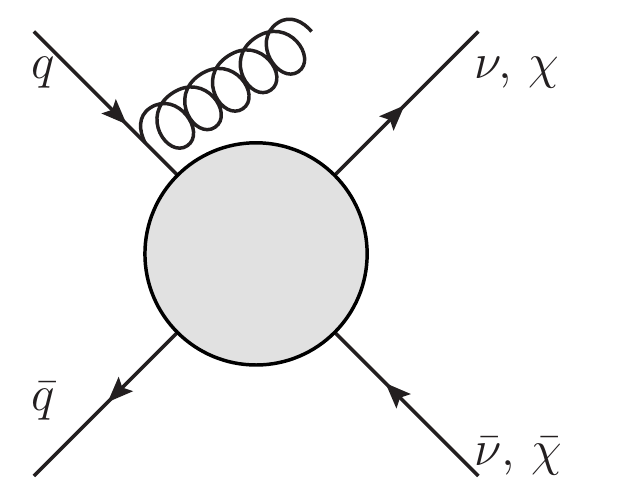} ~~~
  \includegraphics[width=.30\textwidth]{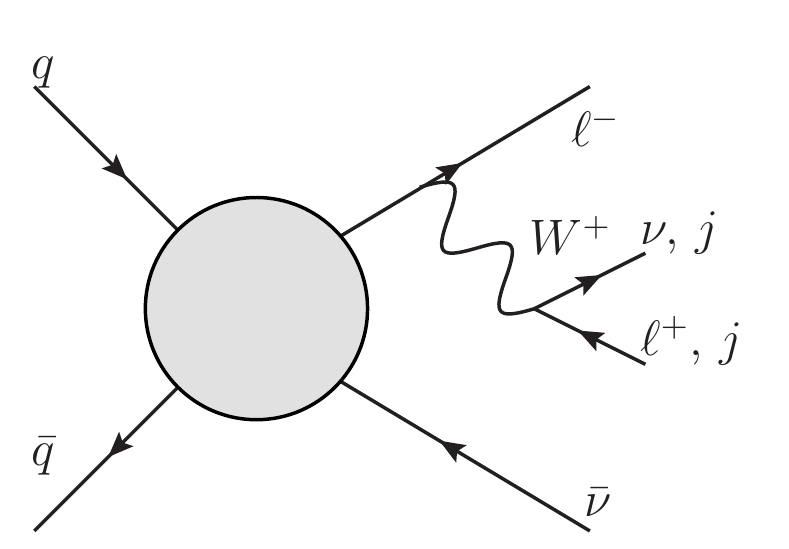} ~~~

\caption{Typical Feynman diagrams for $pp \rightarrow \bar{\nu}\nu (\bar{X}X) + j$ ({\it left panel}) and $pp \rightarrow \overline{\nu}\nu \rightarrow \ell^{\mp} + W^{\pm} + \overline{\nu}$ ({\it right panel}). Though singlet DM and neutrinos are largely degenerate in the former process, only SM neutrinos give rise to the latter process.}
\label{fig:diagrams}
\end{center}
\end{figure*}

Let us first consider whether neutrino magnetic moments below currents limits can produce sizeable missing energy at the LHC?
For Majorana neutrinos the $3 \times3 $ matrix $\mu_{\nu}$ does not have diagonal entries and is anti-symmetric, but is completely general if they are instead Dirac. In the SM the magnetic moment is proportional to the neutrino mass, and therefore extremely small, $\mu_{\nu}^{{\rm SM}} \sim 10^{-20}~\mu_{B}$. 
For Dirac neutrinos in BSM scenarios, naturalness considerations on the coeffecients of effective operators imply, $\mu_{\nu} \lesssim 10^{-14}~\mu_{B}$~\cite{Bell:2005kz}, far below present experimental sensitivity. 
Finally for Majorana neutrinos reactor data as measured by the GEMMA spectrometer constrains, $\mu_{\nu} < 3.2 \times10^{-11}~\mu_{B}$,~\cite{Beda:2010hk} while the 7 TeV LHC sensitivity is around $\sim 3 \times 10^{-5}~{\mu}_{B}$~\cite{Barger:2012pf}, far above what is allowed by reactor and solar data~\cite{Beacom:1999wx}. 
We conclude that neutrino magnetic moments will not produce sizeable missing energy at the LHC.

Proceeding now to operators of mass dimension 6, we turn our attention to the NSI operators between quarks-neutrinos. Non-standard neutrino interactions (NSIs) were first introduced in 1977~\cite{Wolfenstein:1977ue} and continue to be of wide phenomenological interest~\cite{Davidson:2003ha,Friedland:2004pp,Friedland:2004ah,Friedland:2005vy,Scholberg:2005qs,Friedland:2006pi,Kopp:2008ds,Kopp:2007ne,Davidson:2011kr,Friedland:2011za,Friedland:2012tq,Mocioiu:2014gua,Wise:2014oea,Sousa:2015bxa} (see~\cite{Davidson:2003ha,Ohlsson:2012kf} for reviews).

They are constrained by solar~\cite{Friedland:2004pp,Bolanos:2008km,Palazzo:2009rb,Palazzo:2011vg,Bonventre:2013loa,Gonzalez-Garcia:2013usa,Farzan:2015doa,Maltoni:2015kca}, atmospheric~\cite{Fornengo:2001pm,Guzzo:2001mi,GonzalezGarcia:2004wg,Friedland:2004ah,Friedland:2005vy,GonzalezGarcia:2011my,Mocioiu:2014gua}, long-baseline~\cite{Friedland:2006pi,Kopp:2007ne,Kopp:2008ds,GonzalezGarcia:2011my,Friedland:2012tq,Coelho:2012bp,Sousa:2015bxa}, collider~\cite{Berezhiani:2001rs,Davidson:2011kr,Friedland:2011za,Wise:2014oea}, cosmological~\cite{Mangano:2006ar}, and neutrino scattering data~\cite{Davidson:2003ha,Scholberg:2005qs,Ohlsson:2012kf}.

The Lorentz structure of Eq.~\ref{eq:NSI} can be understood as follows. First, assume that NSI can be parameterized as $\mathcal{O}_{NSI} =\mathscr{O}_{\nu}\otimes \mathscr{O}_{f}$ where $\mathscr{O}_{\nu},\mathscr{O}_{f}$ are neutrino and SM fermion bilnears. Under the assumption that lepton number remains a good symmetry and only left-handed neutrinos enter into $\mathscr{O}_{NSI}$, all such operators can be decomposed into $(V -A)\otimes(V -A)$, $(V -A)\otimes(V +A)$ components. 

One may worry that sizeable NSI would also induce large charged lepton interactions~\cite{Bergmann:1999pk,Bergmann:2000gp,Davidson:2003ha}. Indeed, to evade the very strong limits from the charged lepton equivalent of Eq.~(\ref{eq:NSI}) we consider dimension-8 operators of the form~\cite{Berezhiani:2001rs}
\be 
\mathscr{L}_{{\rm NSI}}^{{\rm dim}-8} =-\frac{4 \varepsilon_{\alpha \beta}^{fP}}{v_{EW}^{4}}  \left( \overline{HL_{\alpha}} \gamma_{\mu} HL_{\beta}	\right) \left(\overline{q} \gamma_{\mu} q\right)
\label{eq:dim8}
\ee
where $H$ is the SM Higgs doublet. In unitary gauge and upon electroweak (EW) symmetry breaking we can make the replacement $H \rightarrow \left( h+ v_{EW}\right)/\sqrt{2}$. Thus at low energies, one indeed generates Eq.~(\ref{eq:NSI}) without charged lepton interactions of the same strength. 

The remainder of this paper is organized as follows.  First we introduce our simplified model and calculational framework in Sec.~\ref{sec:model}. In Section~\ref{sec:monojet} we derive new constraints on NSI based on the latest monojet data from the LHC. Then we turn to projections of monojet sensitivity at 13 TeV and the ability to infer the mass of the "missing particles" from the shape of the $\missEt$ distribution. We find that for contact interactions, DM masses $\gtrsim 700$ GeV can be discriminated from NSI  
with about $100~\ifb$ of integrated luminosity. In Sect.~\ref{sec:multilepton} we then use two distinct multi-lepton channels to probe NSI. These searches have neutrino flavor dependent sensitivity and have better sensitivity than monojets for heavy mediators of NSI. In Sec.~\ref{sec:discussion} we discuss the complementarity of these channels along with low-energy probes of NSI for DM-neutrino discrimination and conclude in Sec.~\ref{sec:conclusion}.

\section{Model and Calculational Framework}
\label{sec:model}
In order to derive LHC limits on NSI/DM couplings $\varepsilon$ we have implemented two models in the Universal FeynRules Format (UFO)~\cite{Degrande:2011ua} by adding to the SM  a spin-1 mediator, $R^{\mu}$, which interacts with neutrinos, quarks and DM $X$ through the phenomenological Lagrangians:

\begin{eqnarray}
\mathscr{L}_{{\rm NSI}} &=&g_{\nu} \left(\overline{\nu}P_L\gamma_{\mu}\nu\right)R^{\mu} + 
    \left(\overline{q}\gamma_{\mu}(g_{q}^V+g_q^A\gamma^5)q \right)R^\mu,  \nonumber \\
\mathscr{L}_{{\rm DM}} &=&g_{X} \left(\overline{X}\gamma_{\mu}X\right)R^{\mu} + 
    \left(\overline{q}\gamma_{\mu}(g_{q}^V+g_q^A\gamma^5)q \right)R^\mu,  \nonumber \\
    &+& m_{X} \overline{X} X 
 \label{eq:zprime}
\end{eqnarray}
where $\nu$ and $q$ are summed over all neutrino and quark flavors respectively, and $m_{X}$ is the DM mass.  The Lagrangian $\mathscr{L}_{\rm NSI}$ correctly reproduces the contact interaction, \eq{eq:NSI} when the vector mass, $m_{R}$ is large compared to the center of mass energy. 
{Note that the DM literature tends to report limits on the scale of the dimension-six operator, $\Lambda$, defined as $(\overline{X}\gamma_{\mu}X)(q\gamma^{\mu} q)/\Lambda^{2}$. The conversion from $\Lambda$ to the NSI $\varepsilon$ parameter in this context is, $\varepsilon=(2 G_F\Lambda^2)^{-1}$.}

The main aim of this paper is to illustrate how $\mathscr{L}_{\rm NSI}$ can be discriminated from $\mathscr{L}_{\rm DM}$ 
and gauge the relevant parametric dependencies present in $s$-channel completions of NSI ($t$-channel completions are very strongly constrained~\cite{Friedland:2011za,Wise:2014oea} and not considered further).  For details on a more complete $Z'$ model we refer the reader to the Appendix and to~\cite{Antusch:2008tz,Gavela:2008ra} for additional models.  Furthermore, it is important to highlight that a complete model typically produces signatures {\it in addition} to the monojet and multilepton channels we consider here, making our approach conservative.  

Simplified models of the type in Eq.~\ref{eq:zprime} have been studied extensively in the DM literature~\cite{Bai:2010hh,Graesser:2011vj,Fox:2011pm,Shoemaker:2011vi,Fox:2011qd,Gondolo:2011eq,Lin:2011gj,An:2012va,Frandsen:2012rk,An:2012ue,Alves:2013tqa,Arcadi:2013qia,Lebedev:2014bba,Davidson:2014eia,Fairbairn:2014aqa,Soper:2014ska,Hooper:2014fda,Chala:2015ama}.  While dijet searches provide additional constraints on the models considered here (see e.g.~\cite{An:2012va,Chala:2015ama}) both $\mathscr{L}_{\rm NSI}$ and $\mathscr{L}_{\rm DM}$ contribute equally to this channel and thus it is not a useful discriminatory tool. 

 Our calculational framework is as follows:  To keep the analysis  simple, we consider only vector couplings, i.e. $g_{q}^{A} =g_{X}^{A}= 0$. We import the UFO model into the {\tt MadGraph5\_aMC@NLO} framework
\cite{Alwall:2014hca}, where helicity amplitudes are generated by the {\tt ALOHA}~\cite{deAquino:2011ub} code.  
The hard scattering simulation is then processed through parton showering and hadronization using {\tt Pythia 6}~\cite{Sjostrand:2006za} and {\tt Pythia 8} ~\cite{Sjostrand:2007gs}.
Finally we performed a fast detector simulation for the monojet analysis, using both the PGS~\cite{PGS} and DELPHES~\cite{deFavereau:2013fsa} programs to check our results.

For the monojet computation we use the CTEQ 6L1 \cite{Pumplin:2002vw} set of parton distribution functions,
as this is used by the experimental collaboration, 
and NNPDF 2.3~\cite{Ball:2012cx} for the other processes. We chose the default dynamical factorization and renormalization scales of {\tt MadGraph\_aMC@NLO}. 

\section{Monojet Searches}
\label{sec:monojet}
Any long-lived or stable neutral states, such as neutrinos and DM, with couplings to protons can lead to monojet events at the LHC.  
These monojet processes, depicted in \fig{fig:diagrams} (\emph{left}), are characterized by large missing transverse energy and a very hard jet. 
In \cite{CMS:rwa}, the CMS experiment searched for monojets with $\sqrt{s}=8\TeV$ in the center of mass energy and $\mathcal{L} = 19.5~{\rm fb}^{-1}$ of integrated luminosity, reporting an upper limit at 90\%CL of $\varepsilon=0.053$ for a vector operator
and an invisible particle mass $m_X=1\GeV$. 

\begin{figure}[t]
 \includegraphics[width=.40\textwidth]{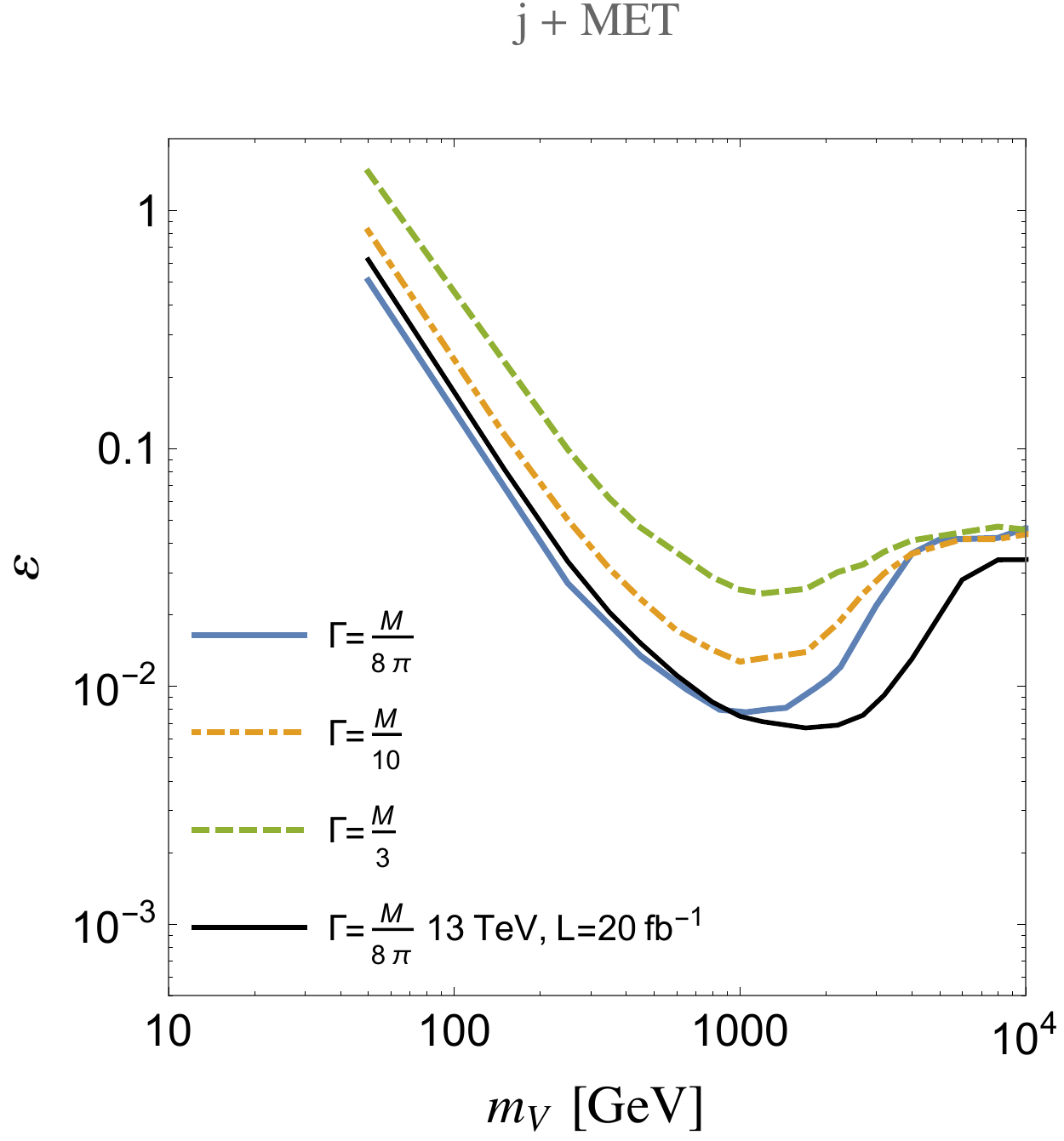} ~~~
\caption{  Here we display the CMS monojet limits~\cite{CMS:rwa} on NSI at 95\%CL for three different choices of the mediator width at $\sqrt{s}=8\TeV$ and with integrated luminosity $\mathcal{L} = 19.5~{\rm fb}^{-1}$. The black solid line denotes the expected limit at 95\%CL  with $\sqrt{s}=13\TeV$ and $L=20 \ifb$. }
\label{fig:monojet}
\end{figure}

To estimate the NSI signal we compute the cross sections for the hard scattering process
\be 
p p \rightarrow V \to \overline{\nu} \nu + 1,\,2\, j ,
\ee
with one and two jets (quarks or gluons), $j$
using the framework described in \sec{sec:model}. 
In particular we use the MLM prescription \cite{Mangano:2006rw} for matching matrix elements with soft jets from the parton shower. 
Following the CMS analysis \cite{CMS:rwa} we require the leading jet to have $p_T(j)>110\GeV$ and to be in the central region of the detector $|\eta(j)|<2.6$. Events with more than 3 jets with $p_T>30\GeV$ and $|\eta|<4.5$ are discarded, while a second jet is allowed as long as the difference in azimuthal angle to the leading jet is less than 2.5, $\Delta\phi(j_1,j_2)<2.5$.  
We further require the missing transverse energy $\missEt>450\GeV$, found to give the best {discriminant}. 

With this analysis set-up we found excellent agreement in the shape of the missing transverse energy distribution for $Z(\nu\nu)+$ jets and $W(\ell\nu)+$ jets SM background. We also found agreement within scale and PDF uncertainties for the number of events. We nevertheless use the fact that the collaboration provides a more precise prediction from data driven techniques and we rescale our predictions by a correction factor of 1.19 to agree with their prediction.

The CMS collaboration report 157 events as the upper 95\% confidence level (CL) limit on the number of events from new physics. Note that a downward fluctuation in the observed number of events gives a constraint about 30$\%$ stronger than expected.  We compute the resulting NSI limits that are shown in \fig{fig:monojet}, as a function of the mediator mass and width ($\Gamma_{V} = \frac{m_{V}}{3},\, \frac{m_{V}}{10},\, \frac{m_{V}}{8\pi}$). 

Note that {\it flavor diagonal} NSIs interfere with the dominant SM background process, $pp \rightarrow Z +j \rightarrow \overline{\nu}\nu +$ jets.  
The strength of the effect depends on the Lorentz structure of the coupling, and the mass of the mediator.  
The effect is small in the contact interaction limit, $\lesssim 5\%$, but can be as large as 20$\%$ when the mass of the mediator is close to the $Z$ mass. 
Although interference is a feature specific to the NSI case it only affects the total number of events and does not aid in distinguishing between dark matter and NSI. We shall therefore omit it in the following 
%

\subsection{Projection to $\sqrt{s}=13\TeV$ LHC and jet $p_T$ shape analysis}

The next LHC run at $\sqrt{s}=13\TeV$ will either further limit or discover NSI and/or DM in monojet searches. In \fig{fig:monojet} we show our projected LHC monojet 95\%CL {limit} as the solid black line for $m_X=0 \GeV$ at $\sqrt{s}=13\TeV$ with the luminosity $\mathcal{L}=20~\ifb$ as expected for the first year of collisions. 
We use the same set-up used at $\sqrt{s}=8\TeV$ and the same normalization rescaling. We assume a systematic error of $5\%$ and performed a $\chi^2$ analysis with which the expected bounds at  $8\TeV$ quoted by CMS were reproduced within error.  
At this luminosity the systematic error dominates and increasing the luminosity further does not appreciably change the experimental sensitivity.

\begin{figure}[t]
 \includegraphics[width=.40\textwidth]{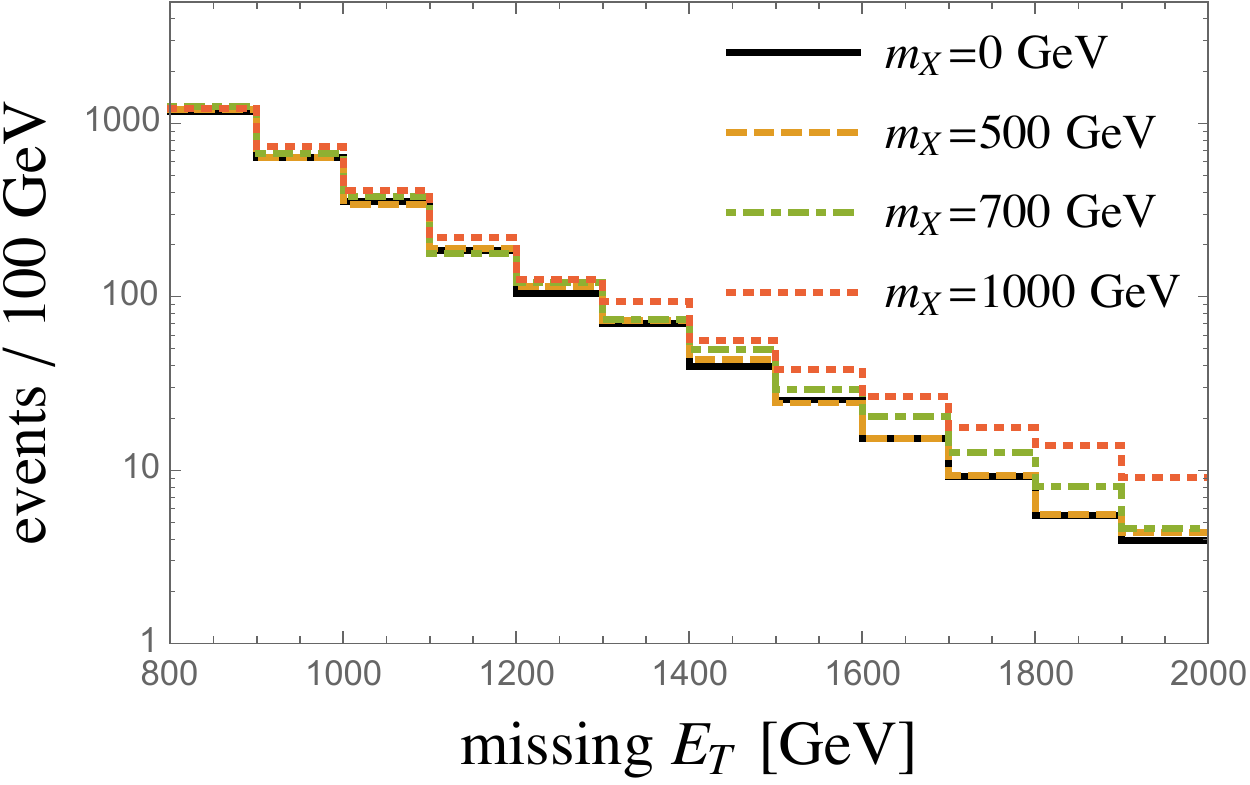} ~~~
\caption{Distribution of events in missing transverse energy, $\missEt$, for $\sqrt{s}=13\TeV$ and $\mathcal{L}=100~\ifb$ for DM masses $m_{X} =$ 0 GeV,~500 GeV,~700 GeV and 1 TeV. Here each distribution is generated assuming contact interactions which produce an identical total number of events, with an interaction strength just below present bounds (see e.g.~Fig.~\ref{fig:monojet}).}
\label{fig:ptshape}
\end{figure}

\begin{figure}[b]
 \includegraphics[width=.40\textwidth]{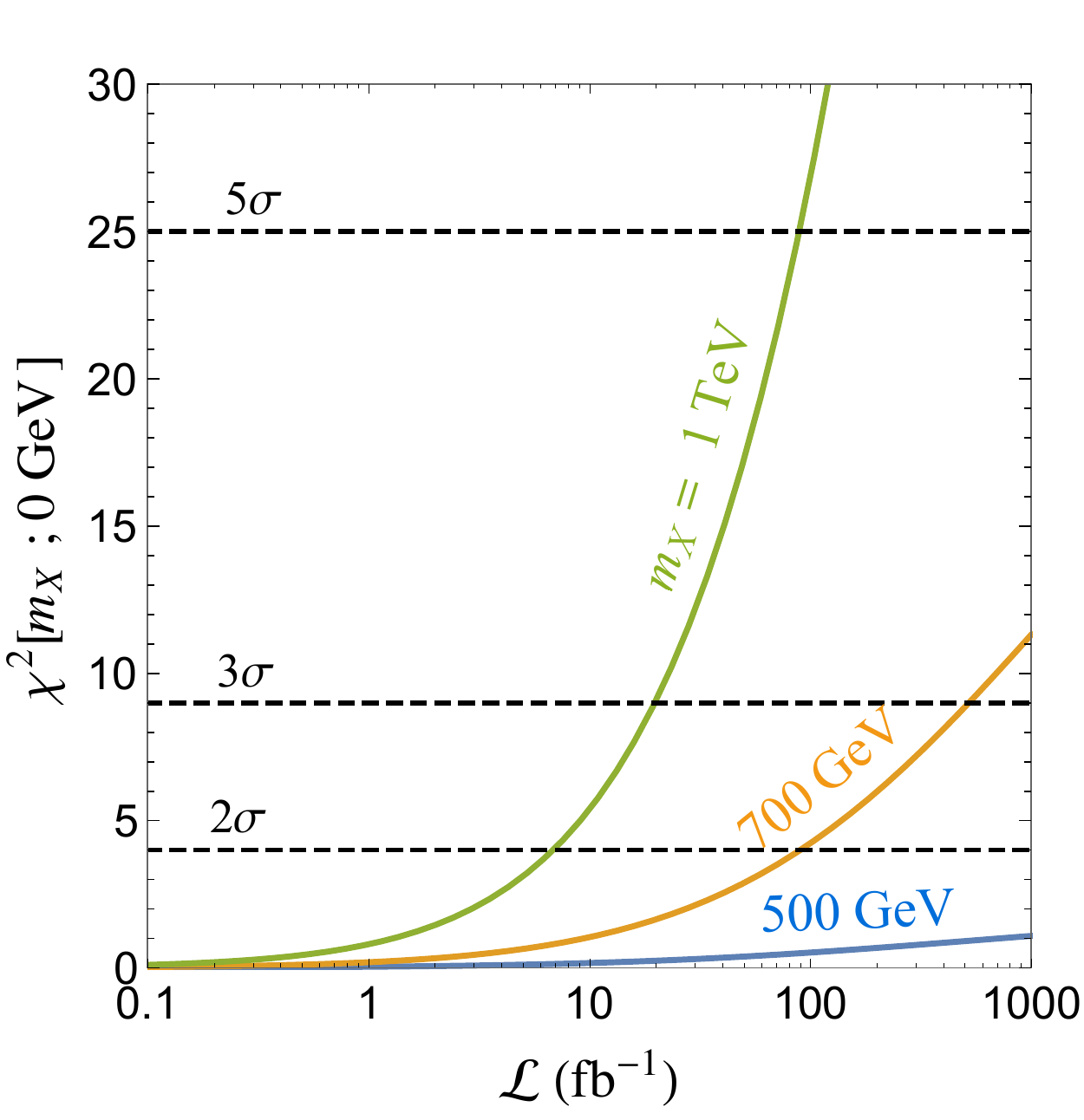} ~~~
\caption{$\chi^2$ projection analysis. Here we generate events at $\sqrt{s}=13\TeV$ in a model with contact interactions just below present bounds with a 0 GeV DM mass. The $\chi^{2}[m_{X} ; 0~{\rm GeV}]$ is computed by fitting $m_{X} = 500, 700, 1000$ GeV DM masses to the input data from a massless invisible particle. The 3$\sigma$, 4$\sigma$ and 5$\sigma$ confidence levels are plotted for reference. }
\label{fig:ptchi}
\end{figure}

The first observable we use to distinguish NSI from DM is the monojet $\missEt$ distribution. Sufficiently heavy DM masses are kinematically relevant at LHC energies and affect the shape of the $\missEt$ distribution.

The allowed value of $\varepsilon$ just below the 95\%CL present limit is $\sim 0.04$ in the contact interaction limit, for massless missing energy particles. 
This situation, \emph{i.e.}  $\varepsilon=0.04$, would produce a $2.9\sigma$ excess according to our projections for the first year Run II of the LHC. 
The same excess of events can be produced for lower $\varepsilon$ but lighter mediator mass or larger $\varepsilon$ and heavier particles in the final state. 
In \fig{fig:ptshape} we show the $\missEt$ distribution for $L=100\ifb$ for DM masses, $m_{X} = 500, 700, 1000$ GeV. The total cross sections are all normalized to the $\varepsilon=0.04$ massless case (shown in the projection \fig{fig:panel})
so that all signals produce the same total number of events with $\missEt > 450$ GeV.  

The shape of the $\missEt$ distributions clearly allow to distinguish between the heaviest and lightest DM masse. We quantify this in a simple $\chi^2$ analysis. In addition to the statistical error, we assume a systematic error per bin of $5\%$ for the background and $20\%$ for the dark matter contribution, as reported in~\cite{CMS:rwa}. The $\chi^2$ distribution is then given by
\be
\chi^2[m_X ; 0~{\rm GeV}] =\sum_i\left[ \frac{S_i(m_X)-S_i(0\GeV)}{\sigma_i} \right]^2 \ , 
\ee
where $S_{i}(m_{X})$ is the number of events in the $i^{\rm th}$ bin. The distribution is shown in \fig{fig:ptchi} for the three masses $m_{X} = 500, 700, 1000$.

\section{MultiLepton Searches}
\label{sec:multilepton}
In addition to the monojet signal, NSI can produce signals in other channels due to the $SU(2)$ charge of neutrinos.  For example, as shown in Fig.~\ref{fig:diagrams} one of the produced neutrinos can radiate a $W$ boson that decays to either jets or $\ell + \nu$,
\be 
p p \rightarrow \overline{\nu} \nu \rightarrow \overline{\nu}+ W^{\pm} \ell^{\mp}.
\ee
Mutli-lepton searches of this type have been used previously to constrain NSI using LHC data~\cite{Davidson:2011kr,Friedland:2011za}.

\begin{figure*}[t]
  \includegraphics[width=.4\textwidth]{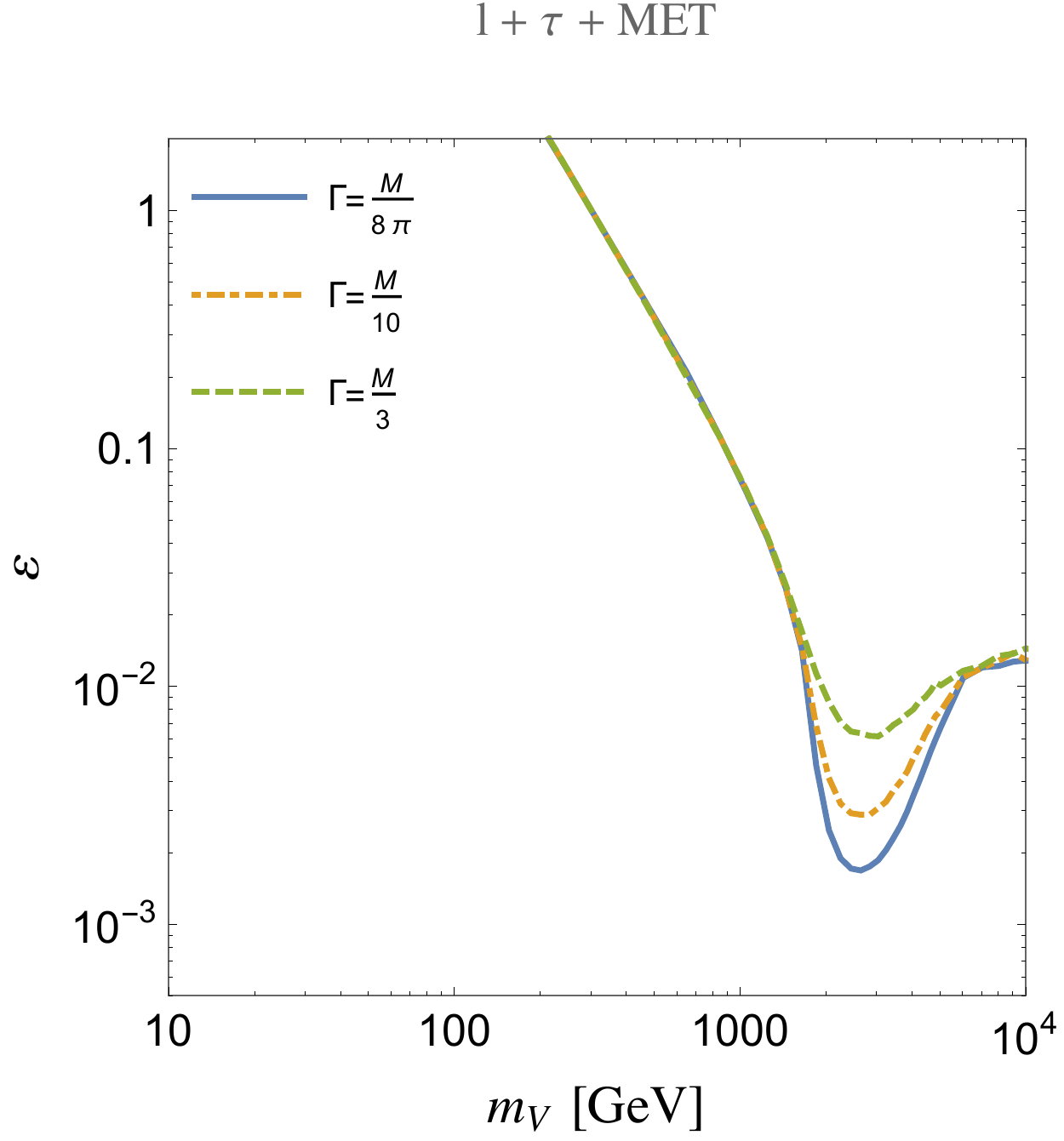}  
    \includegraphics[width=.4\textwidth]{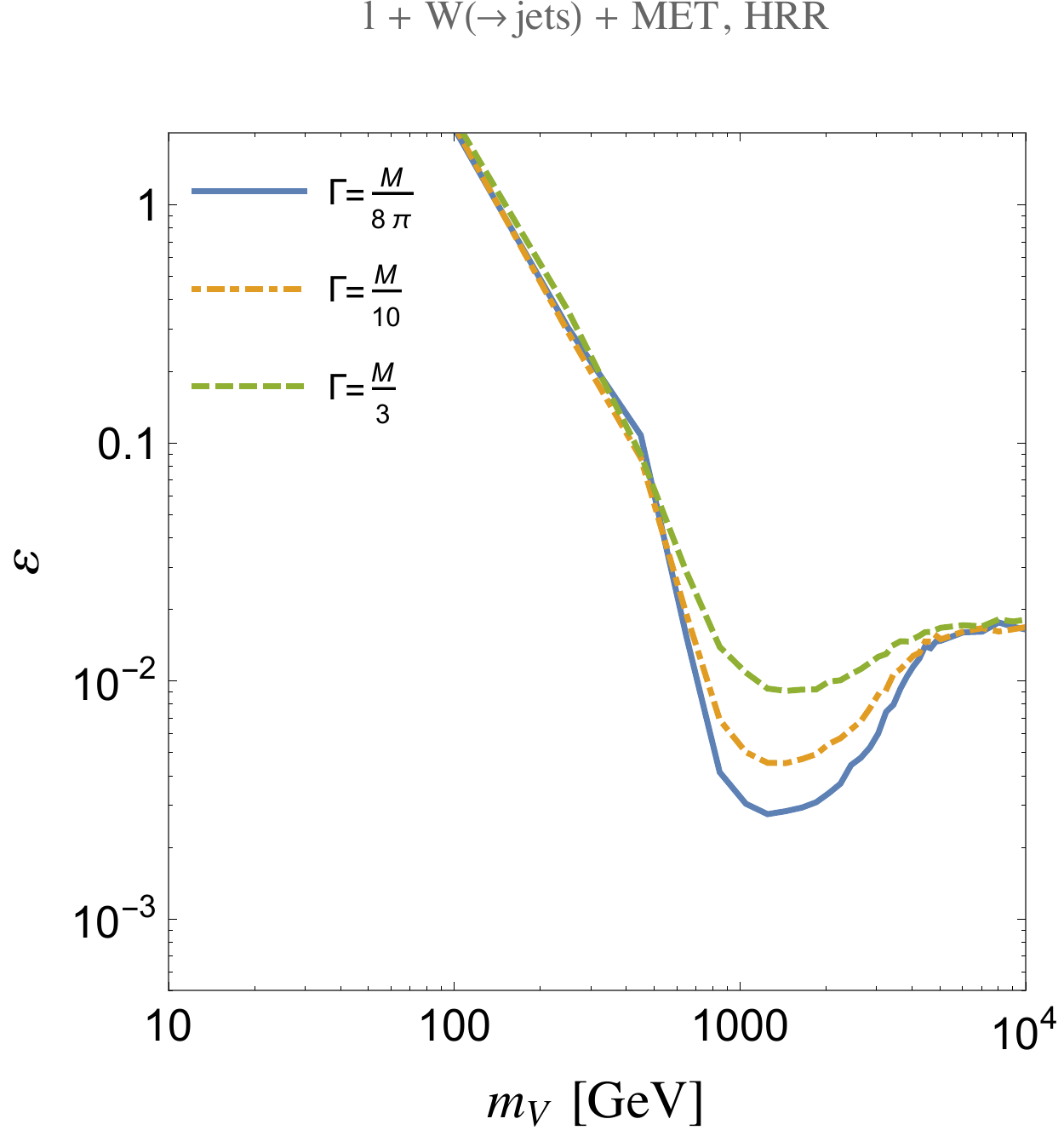}
\caption{ Each panel displaying individual LHC search limits on NSI for three different choices of the mediator width.The {\it left} panel displays the $\tau +\ell + {\rm MET} $ search from ATLAS~\cite{Aad:2014mra}, while the {\it right} panel shows the sensitivity from the $ jj + \ell + {\rm MET}$ search from ATLAS ~\cite{Aad:2015ufa} (see text for details). }
\label{fig:panel}
\end{figure*}

In order to exclude the NSI hypothesis and claim the discovery of a new source of missing energy, we must exclude all possible neutrino flavor structures of NSI. For this it is necessary to consider the lepton in the final state to be a tau, a muon or an electron. Since the mixed flavor interaction, \emph{e.g.} $\varepsilon_{\tau\mu},\,\varepsilon_{e\mu}$, will regardless produce one of these leptons, this condition is also sufficient to constrain mixed terms. 
For the muon and electron in the final state we have relied on the $\sqrt{s} =8$ TeV and $\mathcal{L} = 20.3~{\rm fb}^{-1}$ ATLAS search for resonant diboson production where one boson decays leptonically and the other hadronically~\cite{Aad:2015ufa}. For the tau lepton final state we have used the ATLAS search for supersymmetry with large missing transverse energy, jets and at least one tau lepton, at  $\sqrt{s} =8$ TeV and $\mathcal{L} = 20.3~{\rm fb}^{-1}$ of data~\cite{Aad:2014mra}. We will briefly describe each analysis and results in the following.

The searches we used are not optimized for the NSI signal topologies and we expect that dedicated analyses can improve our results. 
Moreover NSI can lead to signals not considered here but they are expected to be sub-dominant. For example $pp\to \nu\nu (Z\to jj/\ell^+\ell^-)$, where the neutrino radiates a $Z$ boson will suffer from large background from Drell-Yan production. Similarly, in $pp\to \nu\ell (W\to \ell\nu)$ with highly energetic $\ell^+\ell^-$ system, the $W$ cannot be reconstructed, suffering from many more backgrounds. Nonetheless, all these channels may contribute to put bounds on NSI and require a dedicated analysis.

We begin by considering hadronic decays of the $W$s, \emph{i.e}
\subsection{$pp\to\overline{\nu}+ W^{\pm} \ell^{\mp}$, $W^\pm\to jj$, $\ell=e,\,\mu$}

In this analysis 
the $W$-boson is required to be highly boosted {to reduce hadronic backgrounds}. Consequently the two jets from the $W$ are likely to appear as a single jet making jet substructure techniques relevant. The parton-level computation was passed through parton-showering and hadronization using {\tt Pythia 8}. 

We employ the event selection of the experimental analysis in~\cite{Aad:2015ufa}: Leptons are required to have transverse momentum $p_T>25\GeV$ and  $|\eta|<2.5$. Moreover, they are required to satisfy the following isolation criteria: the scalar sum of $p_T$ of tracks with $p_T>1\GeV$ within $\Delta R=\sqrt{\Delta\eta^2+\Delta\phi^2}=0.2$ of the lepton track is required to be less than 15\% of the lepton $p_T$.
The missing transverse energy, defined as the negative of the vectorial sum of the transverse momenta of all electrons, muons and jets within 
$|\eta|<4.9$, is required to be $\missEt>30\GeV$.

We cluster the jets with two different jet definitions provided by {\tt Fastjet 3.1.2}~\cite{Cacciari:2011ma}. For the signal region where the $W$ has large $p_T$ and the jets cluster into a single 'fat' jet, we use the Cambridge algorithm. Otherwise, we use the anti-$k_T$ algorithm with $R=0.4$. 
The ATLAS analysis~\cite{Aad:2015ufa} defines three signal regions as: The merged region (MR), the high-$p_T$ resolved region (HRR) and the low-$p_T$ resolved region (LRR) respectively. In the MR the largest $p_T$ jet $(J)$ is taken to represent a decayed $W$-boson, if it fulfills $p_T(J)>400\GeV$, $|\eta(J)|<2$ and $65\GeV<m(J)<105\GeV$ with the azimuthal angle difference between $J$ and $\vec{\missEt}$ satisfying $\Delta\phi(J,\missEt)<1$. Additionally, the $p_T$ of the lepton and $\missEt$ system is required to be $p_T(\ell\missEt)>400\GeV$. 

If the event does not pass these cuts, we proceed to the resolved region, where the two leading 0.4 anti-$k_T$ jets, $j_{1,2}$ reconstruct a decayed $W$ boson if: $|\eta(j)|<2.8$, $65\GeV<m(jj)<105\GeV$ and $\Delta\phi(j_1,\missEt)<1$. The HRR (LRR) is defined by $p_T(jj)>300(100)\GeV$, $p_T(j)>80(30)\GeV$ and $p_T(\ell\missEt)>300(100)\GeV$. 

After normalizing with an approximate NLO K-factor of K=1.7~\cite{Baur:1994aj}, 
we get reasonable agreement in all three regions for the number of events expected from the SM diboson background. Therefore we assume that this simple analysis is accurate enough for our needs. 

We used the model described by \eq{eq:zprime} to estimate the visible cross sections, $\sigma_S$, and associated number of events, $S=\sigma_S\,\mathcal{L}$ of the NSI signal, where the luminosity is $\mathcal{L}=20.3~\rm{fb}^{-1}$. We rescale our prediction by a K-factor $K=1.2$ to account for QCD corrections extracted from on-shell $Z'$ production \cite{Accomando:2010fz}. 
We moreover assumed a conservative flat theoretical error of $30\%$ to account for PDF and scale uncertainty. The SM prediction for the total number of events and uncertainty, $B\pm\sigma_{B}$, is $161500\pm 2300$, $870\pm40$ and $295\pm22$ for LRR, HRR and MR respectively and the observed number of events, $N^{obs}=157837,\,801$ and $295$ respectively. We summed the errors in quadrature, 
$\sigma_{{\rm TOT}}^2=\sigma_{B}^2+S+(0.3 S)^2$ and estimate the 
95\% CL upper limit on $S$ using a $\chi^2$ analysis, solving for $S$ the equation
\be
\left(\frac{S+B-N^{obs}}{\sigma_{{\rm TOT}}}\right)^{2}=\chi^2_{.05}({\rm d.o.f.}=1)=3.84\,.
\ee
The resulting limits in terms of  $\varepsilon$ are shown in \fig{fig:everything}.

Next we will consider leptonically decaying $W$ bosons 
\subsection{$pp\to\overline{\nu}+ W^{\pm} \tau^{\mp}$, $W^\pm\to \ell\nu$ and 
  $pp\to\overline{\nu}+ W^{\pm} \ell^{\mp}$, $W^\pm\to \tau\nu$, $\tau\to \text{hadrons}$, $\ell=e,\,\mu$}

The signal region defined in the ATLAS search~\cite{Aad:2014mra} relevant for our final state is referred as the $\tau$+lepton ``GMSB signal'' region, which requires a reconstructed hadronically decayed tau lepton and a single isolated electron or muon. Non standard neutrino interactions involving a tau lepton will contribute to this process, but other NSI flavour structures without a tau leptons will equally contribute when the $W$ boson decays to a tau lepton and tau neutrino. 

In our analysis we assume the tau is reconstructed with 70\% of efficiency in this region, as reported in the analysis. In addition, we reproduce the kinematical cuts given therein: $p_T(\ell)>25\GeV$, $p_T(\tau)>20\GeV$, lepton transverse mass, $m_T(\ell)>100\GeV$, defined by 
\be
m_T(\ell) = \sqrt{2 p_T(\ell)\missEt\left(1-\cos(\Delta\phi(\ell,\missEt))\right)}
\ee
and $m_{\rm{eff}}>1700\GeV$, where
\be
m_{\rm{eff}}=p_T(\ell)+p_T(\tau)+\missEt \,.
\ee

The 95\% CL limit on the visible cross section provided by the ATLAS collaboration is 0.20 fb for $\tau+e$ channel and 0.26 fb for the $\tau+\mu$ channel. Using these numbers we find the 95\%CL exclusion limit shown in \fig{fig:everything} as the blue dot-dashed line. 
The limit shown is for NSI involving a tau lepton, $\epsilon_{\tau \tau}$, however the difference with respect to other flavour structures is small.
For $\epsilon_{ee}$ it is only a few percent, and for $\epsilon_{\mu\mu}$ it is about 15\% due to the weaker experimental upper limit on the muon channel.

\begin{figure}[t]
\includegraphics[width=.5\textwidth]{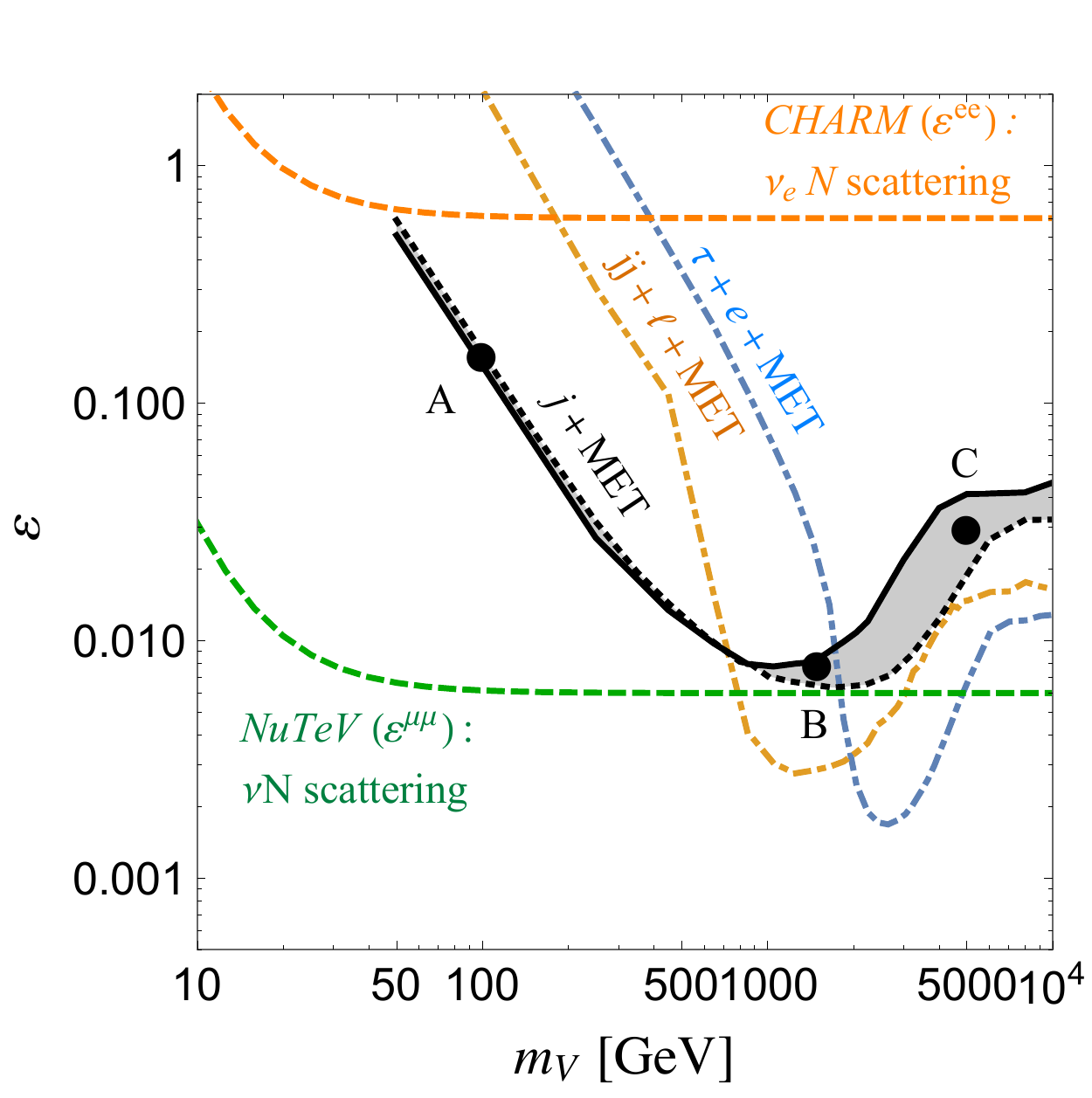} ~~~
\caption{In the mediator mass-coupling plane $(m_{V}, \varepsilon)$, we compare existing searches for NSI from neutrino-nucleus scattering to the LHC mono-jet limits derived in this paper. The upper curve in the gray band depicts the current monojet limits, while the lower curve shows the 13 TeV projection with 100~${\rm fb}^{-1}$. The dot-dashed curves represent the current multi-lepton constraints on NSI based on 8 TeV LHC data.  Additional low-energy constraints on the NSI parameter $\varepsilon_{\alpha \beta}$ include NuTeV's constraint on $\varepsilon_{\mu \mu}$~\cite{Zeller:2001hh} and CHARM's constraint on $\varepsilon_{e e}$~\cite{Dorenbosch:1986tb}. For reference the constraint on $\varepsilon_{\tau \tau}$ is sufficiently weak that it does not appear on the plot (see~\cite{Davidson:2003ha}).}
\label{fig:everything}
\end{figure}

\section{Discussion}
\label{sec:discussion}

After deriving new limits on NSI, we finally asses what future LHC data can unveil. If anomalous missing energy events appear in the next run of the LHC, they will be consistent with either DM or NSI just at the border of the current constraints.
If the events are due to TeV scale DM, then $\missEt$ shape analysis will be enough to rule out NSIs as the origin.
If that is not the case, then we can still use multi-lepton channels to help discriminate between NSIs and DM. To illustrate this point we consider three distinct benchmark scenarios:
\begin{itemize}
\item{\textbf{\textit{Benchmark A}}, $(m_{V},\varepsilon) = (100~{\rm GeV},  0.15).$} \\   The LHC is not a particularly good environment for discriminating neutrinos from DM in the light mediator limit. Although NuTeV's constraint~\cite{Zeller:2001hh} on $\mu$-flavored diagonal NSI {shown as the orange dashed line in Fig.~\ref{fig:everything}} allows us to conclude that this particular flavor structure is not responsible for anomalous missing energy events, the other flavor structures have much weaker constraints and cannot be excluded as potential explanations of LHC monojet signals. NSI with $\tau$ or $e$ flavored interactions can simultaneously escape low-energy probes and multi-lepton searches at the LHC.  {Future data from dedicated low-energy experiments searching for $\nu_{e}-N$ or $\nu_{\tau}-N$ may help resolving this.}

\item{\textbf{\textit{Benchmark B}}, $(m_{V},\varepsilon) = (1500~{\rm GeV},  8\times10^{-3})$.}\\  Here, the LHC's ability to discriminate NSI from the DM hypothesis is much more favorable given the strength of $e,\mu$ flavored NSI limits. Thus although monojet data will be at the discovery level, $\tau$-flavored NSI is degenerate with a DM interpretation. However, since the $\tau + e+{\rm MET}$ search utilized in the present paper, is not optimized for NSI it is possible that a dedicated analysis could help resolve this. 

\item{\textbf{\textit{Benchmark C}}, $(m_{V},\varepsilon) = (5~{\rm TeV},  0.03)$.} \\
 This final benchmark is the most optimistic, as the discrimination between NSI and DM is robust. This is because there are two sensitive probes of NSI since both the $\tau + e + {\rm MET}$ and $jj + \ell + {\rm MET}$ channels yield stronger constraints than monojets. Thus for example, monojet data originating from this benchmark would already be excluded from being of NSI origin with present multi-lepton data. 
 
\end{itemize}

Summarizing, the multi-lepton probes are crucial for distinguishing between DM and NSIs in benchmark C and partially in the case of benchmark B. For benchmark A input from additional low-energy experiments will be needed. These can be either DM or neutrino probes. For example, DM direct detection data can be used to determine the mass of the DM and bound the mediator mass~\cite{Drees:2008bv,McDermott:2011hx,Cherry:2014wia}. 

Alternatively, in the case of neutrinos 
constraints on neutrino scattering will improve shortly. Using the methods outlined in~\cite{Scholberg:2005qs}, the COHERENT~\cite{Bolozdynya:2012xv,Akimov:2013yow} collaboration's multi-target measurement of coherent elastic neutrino-nucleus scattering can be used to substantially strengthen the limits on NSI~\cite{Bolozdynya:2012xv} from NuTeV~\cite{Zeller:2001hh} and CHARM~\cite{Dorenbosch:1986tb}. 

Finally, thanks to the modification of neutrino oscillation probabilities that (vector) NSI induces, long-baseline and solar neutrino data will also further limit NSI. Future probes of NSI include long-baseline experiments such as NO$\nu$A and DUNE~\cite{Friedland:2012tq}, as well as atmospheric data from IceCube DeepCore~\cite{Mocioiu:2014gua}, and solar neutrino data from DM direct detection experiments~\cite{Billard:2014yka}.  

These complementary experimental searches will be tremendously useful in obtaining better sensitivity to NSI.

\section{Conclusion}
\label{sec:conclusion}
If anomalous events with missing energy are found in the next run of the LHC, determining the nature of the ``missing particles'' will be of utmost importance. Given that neutrinos are the only confirmed source of missing energy to date, a neutrino interpretation would be quite natural. Moreover, such non-standard neutrino interactions are rather weakly constrained and could well produce sizeable $j + {\rm MET}$ rates at the LHC.  Here we investigated two useful tools that may aid in this discrimination: $\missEt$ shape analysis of monojet data, and multi-lepton data.  

We found that NSI can be discriminated from DM based on $\missEt$ shape analysis if the DM mass is $\gtrsim 1$ TeV. 

Next, the $SU(2)$ charge of neutrinos implies that NSI contributes in channels involving charged leptons. This gives a simple discriminant between neutrino explanations of missing energy from singlet DM.  To this end we studied $jj + \ell + {\rm MET}$ and $\tau + e + {\rm MET}$ events to derive new limits on NSI. We have found that NSI mediators with masses $\gtrsim 800$ GeV can be fairly robustly discriminated from DM interpretations. This is because the above multi-lepton channels offer greater sensitivity at large mediator masses than monojets. 
In particular, for mediator masses greater than $1.5$ TeV both channels are separately strong enough to discriminate between NSI and DM. 
Light mediator NSI remains hard to probe with LHC data, which underscores the importance of upcoming low-energy probes of NSI.

\acknowledgements
IMS would like to thank the organizers of the {\it Santa Fe Summer Workshop, Implications of Neutrino Flavor Oscillations (INFO) 2015} and the {\it $\nu$@Fermilab} workshop for the opportunity to present this work.  The CP$^3$-Origins center is partially funded by the Danish National Research Foundation, grant number DNRF90.

\section*{Appendix: A UV Model of NSI}
\label{sec:app}
We would like to arrive at a simple completion of Eq.~(\ref{eq:dim8}), which suggests a spin-1 completion given its Lorentz structure.  Moreover, the main feature of Eq.~(\ref{eq:dim8}) is that it implies stronger quark-neutrino interactions than quark-charged lepton interactions.  A simple way to achieve is through the ``baryonic portal''~\cite{Pospelov:2011ha} (though see also~\cite{Pospelov:2012gm,Harnik:2012ni,Pospelov:2013rha,Kopp:2014fha}).  In this class of models the quarks are charged directly under a new $U(1)'$ gauge symmetry.  In addition there are new SM singlet fermions which also carry nonzero $U(1)'$ charge. We will refer to these as ``baryonic neutrinos'' for simplicity. When ordinary and baryonic neutrinos mass mix, the SM neutrinos effectively inherit a small piece of the new interaction.  Thus one needs
\be
\mathcal{L}_{NSI}^{UV} \supset  g_{V}V_{\mu}\left( \overline{Q}\gamma_{\mu}Q +  \overline{\nu_{1}}\gamma_{\mu}\nu_{1}\right) + y NHL + \lambda \phi N \nu_{1}
\ee
where $V$ is the gauge boson of the $U(1)'$ symmetry, $N$ is a singlet fermion, $\nu_{1}$ is the baryonic neutrino, and $\phi$ is a baryonic Higgs whose VEV provides a mass for the $V$. Crucially, once $\phi$ develops a VEV it allows mass mixing with $N$ and hence $N, \nu_{1}$, and $\nu_{\alpha}$ all mass mix. Note that in the above we have used the standard notation for the Lepton doublets, $L_{\alpha} = \left(
\begin{array}{c}
\nu_{\alpha}\\
\ell_{\alpha}\\
\end{array}
\right)$, 
where $\alpha = e, \mu, \tau$,

Next, notice that when $m_{V},m_{N}$ are both large compared to the momentum flowing through the $V$ propagator we can integrate out both new states to write
\be \varepsilon 2 \sqrt{2} G_{F} = \frac{g_{V}^{2}}{m_{V}^{2}}\sin^{2}(2\theta_{b})\,,
\ee
where $\theta_{b}$ is the mixing angle.

An important theoretical constraint on the model comes from anomaly cancellation~\cite{Dobrescu:2014fca}. The least constrained possibility is when the new fermions are vector-like under the SM gauge group, but chiral under $U(1)_{B}$.  Some of these fermion carry electric charge, meaning that they are very strongly constrained. The most conservative constraint comes from chargino searches at LEP and imply that these fermions be heavier than $\sim 90$ GeV~\cite{Heister:2002mn}. 

Since the gauge boson mass is $m_{V} = g_{V} \langle \phi \rangle/2$ and the vector-like fermions have a mass controlled by the new VEV, $m_{f} = \lambda \langle \phi \rangle$, the lower limit on the mass of the fermion translates into an upper limit on the size of the gauge coupling (also assuming $\lambda < \pi$)
\be 
g_{V} = \frac{2\lambda m_{V}}{m_{f}} \lesssim 6.3 \times 10^{-2} \left(\frac{100~{\rm GeV}}{m_{f}}\right)~\left(\frac{m_{V}}{1~\rm{GeV}}\right)
\label{eq:anom}
\ee

Thus the anomaly considerations in Eq. (\ref{eq:anom}) directly constrain the NSI coupling as: 
\be \varepsilon \lesssim 310~\sin^{2}(2 \theta_{b})
\ee
Thus anomaly considerations on their own, are not a significant constraint on NSI models of this sort. However the LHC constraints found in this paper, $\varepsilon \lesssim 10^{-2} -10^{-3}$, can be translated to a constraint on the mixing between the baryonic neutrino $\nu_{1}$ and a SM neutrino, $\sin(2 \theta_{b}) \lesssim (2-6)\times10^{-3} $. 

\bibliographystyle{JHEP}

\bibliography{nu}

\providecommand{\href}[2]{#2}\begingroup\raggedright\begin{thebibliography}{100}

\bibitem{Cirelli:2005uq}
M.~Cirelli, N.~Fornengo, and A.~Strumia, {\it {Minimal dark matter}},  {\em
  Nucl.Phys.} {\bf B753} (2006) 178--194,
  [\href{http://xxx.lanl.gov/abs/hep-ph/0512090}{{\tt hep-ph/0512090}}].

\bibitem{Birkedal:2004xn}
A.~Birkedal, K.~Matchev, and M.~Perelstein, {\it {Dark matter at colliders: A
  Model independent approach}},  {\em Phys. Rev.} {\bf D70} (2004) 077701,
  [\href{http://xxx.lanl.gov/abs/hep-ph/0403004}{{\tt hep-ph/0403004}}].

\bibitem{Cao:2009uw}
Q.-H. Cao, C.-R. Chen, C.~S. Li, and H.~Zhang, {\it {Effective Dark Matter
  Model: Relic density, CDMS II, Fermi LAT and LHC}},  {\em JHEP} {\bf 08}
  (2011) 018, [\href{http://xxx.lanl.gov/abs/0912.4511}{{\tt
  arXiv:0912.4511}}].

\bibitem{Beltran:2010ww}
M.~Beltran, D.~Hooper, E.~W. Kolb, Z.~A.~C. Krusberg, and T.~M.~P. Tait, {\it
  {Maverick dark matter at colliders}},  {\em JHEP} {\bf 09} (2010) 037,
  [\href{http://xxx.lanl.gov/abs/1002.4137}{{\tt arXiv:1002.4137}}].

\bibitem{Goodman:2010yf}
J.~Goodman, M.~Ibe, A.~Rajaraman, W.~Shepherd, T.~M.~P. Tait, and H.-B. Yu,
  {\it {Constraints on Light Majorana dark Matter from Colliders}},  {\em Phys.
  Lett.} {\bf B695} (2011) 185--188,
  [\href{http://xxx.lanl.gov/abs/1005.1286}{{\tt arXiv:1005.1286}}].

\bibitem{Goodman:2010ku}
J.~Goodman, M.~Ibe, A.~Rajaraman, W.~Shepherd, T.~M.~P. Tait, and H.-B. Yu,
  {\it {Constraints on Dark Matter from Colliders}},  {\em Phys. Rev.} {\bf
  D82} (2010) 116010, [\href{http://xxx.lanl.gov/abs/1008.1783}{{\tt
  arXiv:1008.1783}}].

\bibitem{Bai:2010hh}
Y.~Bai, P.~J. Fox, and R.~Harnik, {\it {The Tevatron at the Frontier of Dark
  Matter Direct Detection}},  {\em JHEP} {\bf 12} (2010) 048,
  [\href{http://xxx.lanl.gov/abs/1005.3797}{{\tt arXiv:1005.3797}}].

\bibitem{Fortin:2011hv}
J.-F. Fortin and T.~M.~P. Tait, {\it {Collider Constraints on
  Dipole-Interacting Dark Matter}},  {\em Phys. Rev.} {\bf D85} (2012) 063506,
  [\href{http://xxx.lanl.gov/abs/1103.3289}{{\tt arXiv:1103.3289}}].

\bibitem{Graesser:2011vj}
M.~L. Graesser, I.~M. Shoemaker, and L.~Vecchi, {\it {A Dark Force for
  Baryons}},  \href{http://xxx.lanl.gov/abs/1107.2666}{{\tt arXiv:1107.2666}}.

\bibitem{Fox:2011pm}
P.~J. Fox, R.~Harnik, J.~Kopp, and Y.~Tsai, {\it {Missing Energy Signatures of
  Dark Matter at the LHC}},  {\em Phys. Rev.} {\bf D85} (2012) 056011,
  [\href{http://xxx.lanl.gov/abs/1109.4398}{{\tt arXiv:1109.4398}}].

\bibitem{Friedland:2011za}
A.~Friedland, M.~L. Graesser, I.~M. Shoemaker, and L.~Vecchi, {\it {Probing
  Nonstandard Standard Model Backgrounds with LHC Monojets}},  {\em Phys.
  Lett.} {\bf B714} (2012) 267--275,
  [\href{http://xxx.lanl.gov/abs/1111.5331}{{\tt arXiv:1111.5331}}].

\bibitem{Shoemaker:2011vi}
I.~M. Shoemaker and L.~Vecchi, {\it {Unitarity and Monojet Bounds on Models for
  DAMA, CoGeNT, and CRESST-II}},  {\em Phys.Rev.} {\bf D86} (2012) 015023,
  [\href{http://xxx.lanl.gov/abs/1112.5457}{{\tt arXiv:1112.5457}}].

\bibitem{An:2012va}
H.~An, X.~Ji, and L.-T. Wang, {\it {Light Dark Matter and $Z'$ Dark Force at
  Colliders}},  {\em JHEP} {\bf 07} (2012) 182,
  [\href{http://xxx.lanl.gov/abs/1202.2894}{{\tt arXiv:1202.2894}}].

\bibitem{Fox:2012ee}
P.~J. Fox, R.~Harnik, R.~Primulando, and C.-T. Yu, {\it {Taking a Razor to Dark
  Matter Parameter Space at the LHC}},  {\em Phys. Rev.} {\bf D86} (2012)
  015010, [\href{http://xxx.lanl.gov/abs/1203.1662}{{\tt arXiv:1203.1662}}].

\bibitem{Carpenter:2012rg}
L.~M. Carpenter, A.~Nelson, C.~Shimmin, T.~M. Tait, and D.~Whiteson, {\it
  {Collider searches for dark matter in events with a Z boson and missing
  energy}},  {\em Phys.Rev.} {\bf D87} (2013), no.~7 074005,
  [\href{http://xxx.lanl.gov/abs/1212.3352}{{\tt arXiv:1212.3352}}].

\bibitem{Chatrchyan:2012tea}
{\bf CMS} Collaboration, S.~Chatrchyan et~al., {\it {Search for Dark Matter and
  Large Extra Dimensions in pp Collisions Yielding a Photon and Missing
  Transverse Energy}},  {\em Phys. Rev. Lett.} {\bf 108} (2012) 261803,
  [\href{http://xxx.lanl.gov/abs/1204.0821}{{\tt arXiv:1204.0821}}].

\bibitem{Frandsen:2012rk}
M.~T. Frandsen, F.~Kahlhoefer, A.~Preston, S.~Sarkar, and K.~Schmidt-Hoberg,
  {\it {LHC and Tevatron Bounds on the Dark Matter Direct Detection
  Cross-Section for Vector Mediators}},  {\em JHEP} {\bf 07} (2012) 123,
  [\href{http://xxx.lanl.gov/abs/1204.3839}{{\tt arXiv:1204.3839}}].

\bibitem{Haisch:2012kf}
U.~Haisch, F.~Kahlhoefer, and J.~Unwin, {\it {The impact of heavy-quark loops
  on LHC dark matter searches}},  {\em JHEP} {\bf 07} (2013) 125,
  [\href{http://xxx.lanl.gov/abs/1208.4605}{{\tt arXiv:1208.4605}}].

\bibitem{Bell:2012rg}
N.~F. Bell, J.~B. Dent, A.~J. Galea, T.~D. Jacques, L.~M. Krauss, and T.~J.
  Weiler, {\it {Searching for Dark Matter at the LHC with a Mono-Z}},  {\em
  Phys. Rev.} {\bf D86} (2012) 096011,
  [\href{http://xxx.lanl.gov/abs/1209.0231}{{\tt arXiv:1209.0231}}].

\bibitem{Fox:2012ru}
P.~J. Fox and C.~Williams, {\it {Next-to-Leading Order Predictions for Dark
  Matter Production at Hadron Colliders}},  {\em Phys.Rev.} {\bf D87} (2013),
  no.~5 054030, [\href{http://xxx.lanl.gov/abs/1211.6390}{{\tt
  arXiv:1211.6390}}].

\bibitem{Zhou:2013fla}
N.~Zhou, D.~Berge, and D.~Whiteson, {\it {Mono-everything: combined limits on
  dark matter production at colliders from multiple final states}},  {\em
  Phys.Rev.} {\bf D87} (2013), no.~9 095013,
  [\href{http://xxx.lanl.gov/abs/1302.3619}{{\tt arXiv:1302.3619}}].

\bibitem{Busoni:2013lha}
G.~Busoni, A.~De~Simone, E.~Morgante, and A.~Riotto, {\it {On the Validity of
  the Effective Field Theory for Dark Matter Searches at the LHC}},  {\em
  Phys.Lett.} {\bf B728} (2014) 412--421,
  [\href{http://xxx.lanl.gov/abs/1307.2253}{{\tt arXiv:1307.2253}}].

\bibitem{An:2013xka}
H.~An, L.-T. Wang, and H.~Zhang, {\it {Dark matter with $t$-channel mediator: a
  simple step beyond contact interaction}},  {\em Phys.Rev.} {\bf D89} (2014),
  no.~11 115014, [\href{http://xxx.lanl.gov/abs/1308.0592}{{\tt
  arXiv:1308.0592}}].

\bibitem{Buchmueller:2013dya}
O.~Buchmueller, M.~J. Dolan, and C.~McCabe, {\it {Beyond Effective Field Theory
  for Dark Matter Searches at the LHC}},  {\em JHEP} {\bf 01} (2014) 025,
  [\href{http://xxx.lanl.gov/abs/1308.6799}{{\tt arXiv:1308.6799}}].

\bibitem{Busoni:2014sya}
G.~Busoni, A.~De~Simone, J.~Gramling, E.~Morgante, and A.~Riotto, {\it {On the
  Validity of the Effective Field Theory for Dark Matter Searches at the LHC,
  Part II: Complete Analysis for the $s$-channel}},  {\em JCAP} {\bf 1406}
  (2014) 060, [\href{http://xxx.lanl.gov/abs/1402.1275}{{\tt
  arXiv:1402.1275}}].

\bibitem{Buchmueller:2014yoa}
O.~Buchmueller, M.~J. Dolan, S.~A. Malik, and C.~McCabe, {\it {Characterising
  dark matter searches at colliders and direct detection experiments: Vector
  mediators}},  {\em JHEP} {\bf 01} (2015) 037,
  [\href{http://xxx.lanl.gov/abs/1407.8257}{{\tt arXiv:1407.8257}}].

\bibitem{Abdallah:2014hon}
J.~Abdallah et~al., {\it {Simplified Models for Dark Matter and Missing Energy
  Searches at the LHC}},  \href{http://xxx.lanl.gov/abs/1409.2893}{{\tt
  arXiv:1409.2893}}.

\bibitem{Jacques:2015zha}
T.~Jacques and K.~Nordstr{\"o}m, {\it {Mapping monojet constraints onto
  Simplified Dark Matter Models}},  {\em JHEP} {\bf 06} (2015) 142,
  [\href{http://xxx.lanl.gov/abs/1502.0572}{{\tt arXiv:1502.0572}}].

\bibitem{Chala:2015ama}
M.~Chala, F.~Kahlhoefer, M.~McCullough, G.~Nardini, and K.~Schmidt-Hoberg, {\it
  {Constraining Dark Sectors with Monojets and Dijets}},
  \href{http://xxx.lanl.gov/abs/1503.0591}{{\tt arXiv:1503.0591}}.

\bibitem{Bell:2015sza}
N.~F. Bell, Y.~Cai, J.~B. Dent, R.~K. Leane, and T.~J. Weiler, {\it {Dark
  matter at the LHC: EFTs and gauge invariance}},
  \href{http://xxx.lanl.gov/abs/1503.0787}{{\tt arXiv:1503.0787}}.

\bibitem{Bell:2005kz}
N.~F. Bell, V.~Cirigliano, M.~J. Ramsey-Musolf, P.~Vogel, and M.~B. Wise, {\it
  {How magnetic is the Dirac neutrino?}},  {\em Phys.Rev.Lett.} {\bf 95} (2005)
  151802, [\href{http://xxx.lanl.gov/abs/hep-ph/0504134}{{\tt
  hep-ph/0504134}}].

\bibitem{Beda:2010hk}
A.~Beda, V.~Brudanin, V.~Egorov, D.~Medvedev, V.~Pogosov, et~al., {\it {Upper
  limit on the neutrino magnetic moment from three years of data from the GEMMA
  spectrometer}},  \href{http://xxx.lanl.gov/abs/1005.2736}{{\tt
  arXiv:1005.2736}}.

\bibitem{Barger:2012pf}
V.~Barger, W.-Y. Keung, D.~Marfatia, and P.-Y. Tseng, {\it {Dipole Moment Dark
  Matter at the LHC}},  {\em Phys.Lett.} {\bf B717} (2012) 219--223,
  [\href{http://xxx.lanl.gov/abs/1206.0640}{{\tt arXiv:1206.0640}}].

\bibitem{Beacom:1999wx}
J.~F. Beacom and P.~Vogel, {\it {Neutrino magnetic moments, flavor mixing, and
  the Super-Kamiokande solar data}},  {\em Phys.Rev.Lett.} {\bf 83} (1999)
  5222--5225, [\href{http://xxx.lanl.gov/abs/hep-ph/9907383}{{\tt
  hep-ph/9907383}}].

\bibitem{Wolfenstein:1977ue}
L.~Wolfenstein, {\it {Neutrino Oscillations in Matter}},  {\em Phys.Rev.} {\bf
  D17} (1978) 2369--2374.

\bibitem{Davidson:2003ha}
S.~Davidson, C.~Pena-Garay, N.~Rius, and A.~Santamaria, {\it {Present and
  future bounds on nonstandard neutrino interactions}},  {\em JHEP} {\bf 0303}
  (2003) 011, [\href{http://xxx.lanl.gov/abs/hep-ph/0302093}{{\tt
  hep-ph/0302093}}].

\bibitem{Friedland:2004pp}
A.~Friedland, C.~Lunardini, and C.~Pena-Garay, {\it {Solar neutrinos as probes
  of neutrino matter interactions}},  {\em Phys.Lett.} {\bf B594} (2004) 347,
  [\href{http://xxx.lanl.gov/abs/hep-ph/0402266}{{\tt hep-ph/0402266}}].

\bibitem{Friedland:2004ah}
A.~Friedland, C.~Lunardini, and M.~Maltoni, {\it {Atmospheric neutrinos as
  probes of neutrino-matter interactions}},  {\em Phys.Rev.} {\bf D70} (2004)
  111301, [\href{http://xxx.lanl.gov/abs/hep-ph/0408264}{{\tt
  hep-ph/0408264}}].

\bibitem{Friedland:2005vy}
A.~Friedland and C.~Lunardini, {\it {A Test of tau neutrino interactions with
  atmospheric neutrinos and K2K}},  {\em Phys.Rev.} {\bf D72} (2005) 053009,
  [\href{http://xxx.lanl.gov/abs/hep-ph/0506143}{{\tt hep-ph/0506143}}].

\bibitem{Scholberg:2005qs}
K.~Scholberg, {\it {Prospects for measuring coherent neutrino-nucleus elastic
  scattering at a stopped-pion neutrino source}},  {\em Phys.Rev.} {\bf D73}
  (2006) 033005, [\href{http://xxx.lanl.gov/abs/hep-ex/0511042}{{\tt
  hep-ex/0511042}}].

\bibitem{Friedland:2006pi}
A.~Friedland and C.~Lunardini, {\it {Two modes of searching for new neutrino
  interactions at MINOS}},  {\em Phys.Rev.} {\bf D74} (2006) 033012,
  [\href{http://xxx.lanl.gov/abs/hep-ph/0606101}{{\tt hep-ph/0606101}}].

\bibitem{Kopp:2008ds}
J.~Kopp, T.~Ota, and W.~Winter, {\it {Neutrino factory optimization for
  non-standard interactions}},  {\em Phys.Rev.} {\bf D78} (2008) 053007,
  [\href{http://xxx.lanl.gov/abs/0804.2261}{{\tt arXiv:0804.2261}}].

\bibitem{Kopp:2007ne}
J.~Kopp, M.~Lindner, T.~Ota, and J.~Sato, {\it {Non-standard neutrino
  interactions in reactor and superbeam experiments}},  {\em Phys.Rev.} {\bf
  D77} (2008) 013007, [\href{http://xxx.lanl.gov/abs/0708.0152}{{\tt
  arXiv:0708.0152}}].

\bibitem{Davidson:2011kr}
S.~Davidson and V.~Sanz, {\it {Non-Standard Neutrino Interactions at
  Colliders}},  {\em Phys.Rev.} {\bf D84} (2011) 113011,
  [\href{http://xxx.lanl.gov/abs/1108.5320}{{\tt arXiv:1108.5320}}].

\bibitem{Friedland:2012tq}
A.~Friedland and I.~M. Shoemaker, {\it {Searching for Novel Neutrino
  Interactions at NOvA and Beyond in Light of Large $\theta_{13}$}},
  \href{http://xxx.lanl.gov/abs/1207.6642}{{\tt arXiv:1207.6642}}.

\bibitem{Mocioiu:2014gua}
I.~Mocioiu and W.~Wright, {\it {Non-standard neutrino interactions in the
  mu--tau sector}},  {\em Nucl. Phys.} {\bf B893} (2015) 376--390,
  [\href{http://xxx.lanl.gov/abs/1410.6193}{{\tt arXiv:1410.6193}}].

\bibitem{Wise:2014oea}
M.~B. Wise and Y.~Zhang, {\it {Effective Theory and Simple Completions for
  Neutrino Interactions}},  {\em Phys.Rev.} {\bf D90} (2014), no.~5 053005,
  [\href{http://xxx.lanl.gov/abs/1404.4663}{{\tt arXiv:1404.4663}}].

\bibitem{Sousa:2015bxa}
{\bf MINOS, MINOS+} Collaboration, A.~Sousa, {\it {First MINOS+ Data and New
  Results from MINOS}},  \href{http://xxx.lanl.gov/abs/1502.0771}{{\tt
  arXiv:1502.0771}}.

\bibitem{Ohlsson:2012kf}
T.~Ohlsson, {\it {Status of non-standard neutrino interactions}},  {\em
  Rept.Prog.Phys.} {\bf 76} (2013) 044201,
  [\href{http://xxx.lanl.gov/abs/1209.2710}{{\tt arXiv:1209.2710}}].

\bibitem{Bolanos:2008km}
A.~Bolanos, O.~Miranda, A.~Palazzo, M.~Tortola, and J.~Valle, {\it {Probing
  non-standard neutrino-electron interactions with solar and reactor
  neutrinos}},  {\em Phys.Rev.} {\bf D79} (2009) 113012,
  [\href{http://xxx.lanl.gov/abs/0812.4417}{{\tt arXiv:0812.4417}}].

\bibitem{Palazzo:2009rb}
A.~Palazzo and J.~Valle, {\it {Confusing non-zero $\theta_{13}$ with
  non-standard interactions in the solar neutrino sector}},  {\em Phys.Rev.}
  {\bf D80} (2009) 091301, [\href{http://xxx.lanl.gov/abs/0909.1535}{{\tt
  arXiv:0909.1535}}].

\bibitem{Palazzo:2011vg}
A.~Palazzo, {\it {Hint of non-standard dynamics in solar neutrino conversion}},
   {\em Phys.Rev.} {\bf D83} (2011) 101701,
  [\href{http://xxx.lanl.gov/abs/1101.3875}{{\tt arXiv:1101.3875}}].

\bibitem{Bonventre:2013loa}
R.~Bonventre, A.~LaTorre, J.~Klein, G.~Orebi~Gann, S.~Seibert, et~al., {\it
  {Non-Standard Models, Solar Neutrinos, and Large $\theta_{13}$}},  {\em
  Phys.Rev.} {\bf D88} (2013), no.~5 053010,
  [\href{http://xxx.lanl.gov/abs/1305.5835}{{\tt arXiv:1305.5835}}].

\bibitem{Gonzalez-Garcia:2013usa}
M.~Gonzalez-Garcia and M.~Maltoni, {\it {Determination of matter potential from
  global analysis of neutrino oscillation data}},  {\em JHEP} {\bf 1309} (2013)
  152, [\href{http://xxx.lanl.gov/abs/1307.3092}{{\tt arXiv:1307.3092}}].

\bibitem{Farzan:2015doa}
Y.~Farzan, {\it {A model for large non-standard interactions of neutrinos
  leading to the LMA-Dark solution}},
  \href{http://xxx.lanl.gov/abs/1505.0690}{{\tt arXiv:1505.0690}}.

\bibitem{Maltoni:2015kca}
M.~Maltoni and A.~{\relax Yu}. Smirnov, {\it {Solar neutrinos and neutrino
  physics}},  \href{http://xxx.lanl.gov/abs/1507.0528}{{\tt arXiv:1507.0528}}.

\bibitem{Fornengo:2001pm}
N.~Fornengo, M.~Maltoni, R.~Tomas, and J.~Valle, {\it {Probing neutrino
  nonstandard interactions with atmospheric neutrino data}},  {\em Phys.Rev.}
  {\bf D65} (2002) 013010, [\href{http://xxx.lanl.gov/abs/hep-ph/0108043}{{\tt
  hep-ph/0108043}}].

\bibitem{Guzzo:2001mi}
M.~Guzzo, P.~de~Holanda, M.~Maltoni, H.~Nunokawa, M.~Tortola, et~al., {\it
  {Status of a hybrid three neutrino interpretation of neutrino data}},  {\em
  Nucl.Phys.} {\bf B629} (2002) 479--490,
  [\href{http://xxx.lanl.gov/abs/hep-ph/0112310}{{\tt hep-ph/0112310}}].

\bibitem{GonzalezGarcia:2004wg}
M.~Gonzalez-Garcia and M.~Maltoni, {\it {Atmospheric neutrino oscillations and
  new physics}},  {\em Phys.Rev.} {\bf D70} (2004) 033010,
  [\href{http://xxx.lanl.gov/abs/hep-ph/0404085}{{\tt hep-ph/0404085}}].

\bibitem{GonzalezGarcia:2011my}
M.~Gonzalez-Garcia, M.~Maltoni, and J.~Salvado, {\it {Testing matter effects in
  propagation of atmospheric and long-baseline neutrinos}},  {\em JHEP} {\bf
  1105} (2011) 075, [\href{http://xxx.lanl.gov/abs/1103.4365}{{\tt
  arXiv:1103.4365}}].

\bibitem{Coelho:2012bp}
J.~A.~B. Coelho, T.~Kafka, W.~A. Mann, J.~Schneps, and O.~Altinok, {\it
  {Constraints for non-standard interaction $\epsilon\_{e \tau}V\_e$ from
  $\nu\_e$ appearance in MINOS and T2K}},  {\em Phys. Rev.} {\bf D86} (2012)
  113015, [\href{http://xxx.lanl.gov/abs/1209.3757}{{\tt arXiv:1209.3757}}].

\bibitem{Berezhiani:2001rs}
Z.~Berezhiani and A.~Rossi, {\it {Limits on the nonstandard interactions of
  neutrinos from e+ e- colliders}},  {\em Phys.Lett.} {\bf B535} (2002)
  207--218, [\href{http://xxx.lanl.gov/abs/hep-ph/0111137}{{\tt
  hep-ph/0111137}}].

\bibitem{Mangano:2006ar}
G.~Mangano, G.~Miele, S.~Pastor, T.~Pinto, O.~Pisanti, et~al., {\it {Effects of
  non-standard neutrino-electron interactions on relic neutrino decoupling}},
  {\em Nucl.Phys.} {\bf B756} (2006) 100--116,
  [\href{http://xxx.lanl.gov/abs/hep-ph/0607267}{{\tt hep-ph/0607267}}].

\bibitem{Bergmann:1999pk}
S.~Bergmann, Y.~Grossman, and D.~M. Pierce, {\it {Can lepton flavor violating
  interactions explain the atmospheric neutrino problem?}},  {\em Phys.Rev.}
  {\bf D61} (2000) 053005, [\href{http://xxx.lanl.gov/abs/hep-ph/9909390}{{\tt
  hep-ph/9909390}}].

\bibitem{Bergmann:2000gp}
S.~Bergmann, M.~Guzzo, P.~de~Holanda, P.~Krastev, and H.~Nunokawa, {\it {Status
  of the solution to the solar neutrino problem based on nonstandard neutrino
  interactions}},  {\em Phys.Rev.} {\bf D62} (2000) 073001,
  [\href{http://xxx.lanl.gov/abs/hep-ph/0004049}{{\tt hep-ph/0004049}}].

\bibitem{Degrande:2011ua}
C.~Degrande, C.~Duhr, B.~Fuks, D.~Grellscheid, O.~Mattelaer, et~al., {\it {UFO
  - The Universal FeynRules Output}},  {\em Comput.Phys.Commun.} {\bf 183}
  (2012) 1201--1214, [\href{http://xxx.lanl.gov/abs/1108.2040}{{\tt
  arXiv:1108.2040}}].

\bibitem{Antusch:2008tz}
S.~Antusch, J.~P. Baumann, and E.~Fernandez-Martinez, {\it {Non-Standard
  Neutrino Interactions with Matter from Physics Beyond the Standard Model}},
  {\em Nucl.Phys.} {\bf B810} (2009) 369--388,
  [\href{http://xxx.lanl.gov/abs/0807.1003}{{\tt arXiv:0807.1003}}].

\bibitem{Gavela:2008ra}
M.~Gavela, D.~Hernandez, T.~Ota, and W.~Winter, {\it {Large gauge invariant
  non-standard neutrino interactions}},  {\em Phys.Rev.} {\bf D79} (2009)
  013007, [\href{http://xxx.lanl.gov/abs/0809.3451}{{\tt arXiv:0809.3451}}].

\bibitem{Fox:2011qd}
P.~J. Fox, J.~Liu, D.~Tucker-Smith, and N.~Weiner, {\it {An Effective Z'}},
  {\em Phys.Rev.} {\bf D84} (2011) 115006,
  [\href{http://xxx.lanl.gov/abs/1104.4127}{{\tt arXiv:1104.4127}}].

\bibitem{Gondolo:2011eq}
P.~Gondolo, P.~Ko, and Y.~Omura, {\it {Light dark matter in leptophobic Z'
  models}},  {\em Phys.Rev.} {\bf D85} (2012) 035022,
  [\href{http://xxx.lanl.gov/abs/1106.0885}{{\tt arXiv:1106.0885}}].

\bibitem{Lin:2011gj}
T.~Lin, H.-B. Yu, and K.~M. Zurek, {\it {On Symmetric and Asymmetric Light Dark
  Matter}},  {\em Phys.Rev.} {\bf D85} (2012) 063503,
  [\href{http://xxx.lanl.gov/abs/1111.0293}{{\tt arXiv:1111.0293}}].

\bibitem{An:2012ue}
H.~An, R.~Huo, and L.-T. Wang, {\it {Searching for Low Mass Dark Portal at the
  LHC}},  {\em Phys.Dark Univ.} {\bf 2} (2013) 50--57,
  [\href{http://xxx.lanl.gov/abs/1212.2221}{{\tt arXiv:1212.2221}}].

\bibitem{Alves:2013tqa}
A.~Alves, S.~Profumo, and F.~S. Queiroz, {\it {The dark $Z^{'}$ portal: direct,
  indirect and collider searches}},  {\em JHEP} {\bf 1404} (2014) 063,
  [\href{http://xxx.lanl.gov/abs/1312.5281}{{\tt arXiv:1312.5281}}].

\bibitem{Arcadi:2013qia}
G.~Arcadi, Y.~Mambrini, M.~H.~G. Tytgat, and B.~Zaldivar, {\it {Invisible
  $Z^\prime$ and dark matter: LHC vs LUX constraints}},  {\em JHEP} {\bf 1403}
  (2014) 134, [\href{http://xxx.lanl.gov/abs/1401.0221}{{\tt
  arXiv:1401.0221}}].

\bibitem{Lebedev:2014bba}
O.~Lebedev and Y.~Mambrini, {\it {Axial dark matter: The case for an invisible
  $Z′$}},  {\em Phys.Lett.} {\bf B734} (2014) 350--353,
  [\href{http://xxx.lanl.gov/abs/1403.4837}{{\tt arXiv:1403.4837}}].

\bibitem{Davidson:2014eia}
S.~Davidson, {\it {Including the Z in an Effective Field Theory for dark matter
  at the LHC}},  {\em JHEP} {\bf 1410} (2014) 84,
  [\href{http://xxx.lanl.gov/abs/1403.5161}{{\tt arXiv:1403.5161}}].

\bibitem{Fairbairn:2014aqa}
M.~Fairbairn and J.~Heal, {\it {Complementarity of dark matter searches at
  resonance}},  {\em Phys.Rev.} {\bf D90} (2014), no.~11 115019,
  [\href{http://xxx.lanl.gov/abs/1406.3288}{{\tt arXiv:1406.3288}}].

\bibitem{Soper:2014ska}
D.~E. Soper, M.~Spannowsky, C.~J. Wallace, and T.~M.~P. Tait, {\it {Scattering
  of Dark Particles with Light Mediators}},  {\em Phys.Rev.} {\bf D90} (2014),
  no.~11 115005, [\href{http://xxx.lanl.gov/abs/1407.2623}{{\tt
  arXiv:1407.2623}}].

\bibitem{Hooper:2014fda}
D.~Hooper, {\it {$Z′$ mediated dark matter models for the Galactic Center
  gamma-ray excess}},  {\em Phys.Rev.} {\bf D91} (2015) 035025,
  [\href{http://xxx.lanl.gov/abs/1411.4079}{{\tt arXiv:1411.4079}}].

\bibitem{Alwall:2014hca}
J.~Alwall, R.~Frederix, S.~Frixione, V.~Hirschi, F.~Maltoni, et~al., {\it {The
  automated computation of tree-level and next-to-leading order differential
  cross sections, and their matching to parton shower simulations}},  {\em
  JHEP} {\bf 1407} (2014) 079, [\href{http://xxx.lanl.gov/abs/1405.0301}{{\tt
  arXiv:1405.0301}}].

\bibitem{deAquino:2011ub}
P.~de~Aquino, W.~Link, F.~Maltoni, O.~Mattelaer, and T.~Stelzer, {\it {ALOHA:
  Automatic Libraries Of Helicity Amplitudes for Feynman Diagram
  Computations}},  {\em Comput.Phys.Commun.} {\bf 183} (2012) 2254--2263,
  [\href{http://xxx.lanl.gov/abs/1108.2041}{{\tt arXiv:1108.2041}}].

\bibitem{Sjostrand:2006za}
T.~Sjostrand, S.~Mrenna, and P.~Z. Skands, {\it {PYTHIA 6.4 Physics and
  Manual}},  {\em JHEP} {\bf 05} (2006) 026,
  [\href{http://xxx.lanl.gov/abs/hep-ph/0603175}{{\tt hep-ph/0603175}}].

\bibitem{Sjostrand:2007gs}
T.~Sjostrand, S.~Mrenna, and P.~Z. Skands, {\it {A Brief Introduction to PYTHIA
  8.1}},  {\em Comput.Phys.Commun.} {\bf 178} (2008) 852--867,
  [\href{http://xxx.lanl.gov/abs/0710.3820}{{\tt arXiv:0710.3820}}].

\bibitem{PGS}
J.~C. et~al., {\it {PGS-Pretty Good Simulation of high energy collision}},
  \href{http://xxx.lanl.gov/abs/http://www.physics.ucdavis.edu/~conway/research/software/pgs/pgs4-general.htm}{{\tt
  http://www.physics.ucdavis.edu/~conway/research/software/pgs/pgs4-general.htm}}.

\bibitem{deFavereau:2013fsa}
{\bf DELPHES 3} Collaboration, J.~de~Favereau et~al., {\it {DELPHES 3, A
  modular framework for fast simulation of a generic collider experiment}},
  {\em JHEP} {\bf 1402} (2014) 057,
  [\href{http://xxx.lanl.gov/abs/1307.6346}{{\tt arXiv:1307.6346}}].

\bibitem{Pumplin:2002vw}
J.~Pumplin, D.~R. Stump, J.~Huston, H.~L. Lai, P.~M. Nadolsky, and W.~K. Tung,
  {\it {New generation of parton distributions with uncertainties from global
  QCD analysis}},  {\em JHEP} {\bf 07} (2002) 012,
  [\href{http://xxx.lanl.gov/abs/hep-ph/0201195}{{\tt hep-ph/0201195}}].

\bibitem{Ball:2012cx}
R.~D. Ball, V.~Bertone, S.~Carrazza, C.~S. Deans, L.~Del~Debbio, et~al., {\it
  {Parton distributions with LHC data}},  {\em Nucl.Phys.} {\bf B867} (2013)
  244--289, [\href{http://xxx.lanl.gov/abs/1207.1303}{{\tt arXiv:1207.1303}}].

\bibitem{CMS:rwa}
{\bf CMS} Collaboration, {\it {Search for new physics in monojet events in pp
  collisions at sqrt(s)= 8 TeV}}, .

\bibitem{Mangano:2006rw}
M.~L. Mangano, M.~Moretti, F.~Piccinini, and M.~Treccani, {\it {Matching matrix
  elements and shower evolution for top-quark production in hadronic
  collisions}},  {\em JHEP} {\bf 01} (2007) 013,
  [\href{http://xxx.lanl.gov/abs/hep-ph/0611129}{{\tt hep-ph/0611129}}].

\bibitem{Aad:2014mra}
{\bf ATLAS} Collaboration, G.~Aad et~al., {\it {Search for supersymmetry in
  events with large missing transverse momentum, jets, and at least one tau
  lepton in 20 fb$^{-1}$ of $\sqrt{s}=$ 8 TeV proton-proton collision data with
  the ATLAS detector}},  {\em JHEP} {\bf 1409} (2014) 103,
  [\href{http://xxx.lanl.gov/abs/1407.0603}{{\tt arXiv:1407.0603}}].

\bibitem{Aad:2015ufa}
{\bf ATLAS} Collaboration, G.~Aad et~al., {\it {Search for production of
  $WW/WZ$ resonances decaying to a lepton, neutrino and jets in $pp$ collisions
  at $\sqrt{s}$ = 8 TeV with the ATLAS detector}},
  \href{http://xxx.lanl.gov/abs/1503.0467}{{\tt arXiv:1503.0467}}.

\bibitem{Cacciari:2011ma}
M.~Cacciari, G.~P. Salam, and G.~Soyez, {\it {FastJet User Manual}},  {\em
  Eur.Phys.J.} {\bf C72} (2012) 1896,
  [\href{http://xxx.lanl.gov/abs/1111.6097}{{\tt arXiv:1111.6097}}].

\bibitem{Baur:1994aj}
U.~Baur, T.~Han, and J.~Ohnemus, {\it {$W Z$ production at hadron colliders:
  Effects of nonstandard $W W Z$ couplings and QCD corrections}},  {\em Phys.
  Rev.} {\bf D51} (1995) 3381--3407,
  [\href{http://xxx.lanl.gov/abs/hep-ph/9410266}{{\tt hep-ph/9410266}}].

\bibitem{Accomando:2010fz}
E.~Accomando, A.~Belyaev, L.~Fedeli, S.~F. King, and
  C.~Shepherd-Themistocleous, {\it {Z' physics with early LHC data}},  {\em
  Phys. Rev.} {\bf D83} (2011) 075012,
  [\href{http://xxx.lanl.gov/abs/1010.6058}{{\tt arXiv:1010.6058}}].

\bibitem{Zeller:2001hh}
{\bf NuTeV} Collaboration, G.~P. Zeller et~al., {\it {A Precise determination
  of electroweak parameters in neutrino nucleon scattering}},  {\em Phys. Rev.
  Lett.} {\bf 88} (2002) 091802,
  [\href{http://xxx.lanl.gov/abs/hep-ex/0110059}{{\tt hep-ex/0110059}}].
  [Erratum: Phys. Rev. Lett.90,239902(2003)].

\bibitem{Dorenbosch:1986tb}
{\bf CHARM} Collaboration, J.~Dorenbosch et~al., {\it {Experimental
  Verification of the Universality of $\nu\_e$ and $\nu\_\mu$ Coupling to the
  Neutral Weak Current}},  {\em Phys. Lett.} {\bf B180} (1986) 303.

\bibitem{Drees:2008bv}
M.~Drees and C.-L. Shan, {\it {Model-Independent Determination of the WIMP Mass
  from Direct Dark Matter Detection Data}},  {\em JCAP} {\bf 0806} (2008) 012,
  [\href{http://xxx.lanl.gov/abs/0803.4477}{{\tt arXiv:0803.4477}}].

\bibitem{McDermott:2011hx}
S.~D. McDermott, H.-B. Yu, and K.~M. Zurek, {\it {The Dark Matter Inverse
  Problem: Extracting Particle Physics from Scattering Events}},  {\em
  Phys.Rev.} {\bf D85} (2012) 123507,
  [\href{http://xxx.lanl.gov/abs/1110.4281}{{\tt arXiv:1110.4281}}].

\bibitem{Cherry:2014wia}
J.~F. Cherry, M.~T. Frandsen, and I.~M. Shoemaker, {\it {Halo Independent
  Direct Detection of Momentum-Dependent Dark Matter}},  {\em JCAP} {\bf 1410}
  (2014), no.~10 022, [\href{http://xxx.lanl.gov/abs/1405.1420}{{\tt
  arXiv:1405.1420}}].

\bibitem{Bolozdynya:2012xv}
A.~Bolozdynya, F.~Cavanna, Y.~Efremenko, G.~Garvey, V.~Gudkov, et~al., {\it
  {Opportunities for Neutrino Physics at the Spallation Neutron Source: A White
  Paper}},  \href{http://xxx.lanl.gov/abs/1211.5199}{{\tt arXiv:1211.5199}}.

\bibitem{Akimov:2013yow}
{\bf CSI} Collaboration, D.~Akimov et~al., {\it {Coherent Scattering
  Investigations at the Spallation Neutron Source: a Snowmass White Paper}},
  in {\em {Community Summer Study 2013: Snowmass on the Mississippi (CSS2013)
  Minneapolis, MN, USA, July 29-August 6, 2013}}, 2013.
\newblock \href{http://xxx.lanl.gov/abs/1310.0125}{{\tt arXiv:1310.0125}}.

\bibitem{Billard:2014yka}
J.~Billard, L.~Strigari, and E.~Figueroa-Feliciano, {\it {Solar neutrino
  physics with low-threshold dark matter detectors}},  {\em Phys. Rev.} {\bf
  D91} (2015), no.~9 095023, [\href{http://xxx.lanl.gov/abs/1409.0050}{{\tt
  arXiv:1409.0050}}].

\bibitem{Pospelov:2011ha}
M.~Pospelov, {\it {Neutrino Physics with Dark Matter Experiments and the
  Signature of New Baryonic Neutral Currents}},  {\em Phys.Rev.} {\bf D84}
  (2011) 085008, [\href{http://xxx.lanl.gov/abs/1103.3261}{{\tt
  arXiv:1103.3261}}].

\bibitem{Pospelov:2012gm}
M.~Pospelov and J.~Pradler, {\it {Elastic scattering signals of solar neutrinos
  with enhanced baryonic currents}},  {\em Phys.Rev.} {\bf D85} (2012) 113016,
  [\href{http://xxx.lanl.gov/abs/1203.0545}{{\tt arXiv:1203.0545}}].

\bibitem{Harnik:2012ni}
R.~Harnik, J.~Kopp, and P.~A. Machado, {\it {Exploring nu Signals in Dark
  Matter Detectors}},  {\em JCAP} {\bf 1207} (2012) 026,
  [\href{http://xxx.lanl.gov/abs/1202.6073}{{\tt arXiv:1202.6073}}].

\bibitem{Pospelov:2013rha}
M.~Pospelov and J.~Pradler, {\it {Dark Matter or Neutrino recoil?
  Interpretation of Recent Experimental Results}},  {\em Phys.Rev.} {\bf D89}
  (2014) 055012, [\href{http://xxx.lanl.gov/abs/1311.5764}{{\tt
  arXiv:1311.5764}}].

\bibitem{Kopp:2014fha}
J.~Kopp and J.~Welter, {\it {The Not-So-Sterile 4th Neutrino: Constraints on
  New Gauge Interactions from Neutrino Oscillation Experiments}},  {\em JHEP}
  {\bf 1412} (2014) 104, [\href{http://xxx.lanl.gov/abs/1408.0289}{{\tt
  arXiv:1408.0289}}].

\bibitem{Dobrescu:2014fca}
B.~A. Dobrescu and C.~Frugiuele, {\it {Hidden GeV-scale interactions of
  quarks}},  {\em Phys.Rev.Lett.} {\bf 113} (2014) 061801,
  [\href{http://xxx.lanl.gov/abs/1404.3947}{{\tt arXiv:1404.3947}}].

\bibitem{Heister:2002mn}
{\bf ALEPH} Collaboration, A.~Heister et~al., {\it {Search for charginos nearly
  mass degenerate with the lightest neutralino in e+ e- collisions at
  center-of-mass energies up to 209-GeV}},  {\em Phys. Lett.} {\bf B533} (2002)
  223--236, [\href{http://xxx.lanl.gov/abs/hep-ex/0203020}{{\tt
  hep-ex/0203020}}].

\end{thebibliography}\endgroup

\end{document}